%% file: withoutrev_proposal.tex
\documentclass[12pt]{article}

\newif\ifpdf
\ifx\pdfoutput\undefined
    \pdffalse           
\else
    \pdfoutput=1        
    \pdftrue
\fi

\usepackage{graphicx,picinpar}               

\ifpdf
\usepackage[pdftex]{hyperref}
\fi

\newcommand{\jeanfig}[3]{\begin{figure}\begin{center}
\includegraphics[height=#2in]{Graphics/#1}\end{center}
\caption{\footnotesize #3\label{f:#1}}\end{figure}}

\begin{document}
\begin{titlepage}
\begin{flushright}
UPR-0241E
\end{flushright}

\vspace*{1.2in}

\noindent{\Large \bf Megaton Modular Multi-Purpose Neutrino Detector

\vspace{1ex}

\hspace*{2.3in} for a 

\vspace{1.5ex}

\hspace*{.13in} Program of Physics in the Homestake DUSEL}

\vspace{1.5in}

\noindent The contributors to this document are aware that an undertaking of
the magnitude proposed here will require significantly more collaborators than are listed below. Accordingly, they invite and welcome individuals with similar interests in any part of this enterprise to join with them by informing any of the undersigned.

\vspace{.5in}

\noindent A meeting of the interested parties will be held soon.

\vspace{.7in}

\begin{tabular}{ll}

BNL: & M.V. Diwan, R.L. Hahn, W. Marciano, B. Viren\\
LSU: & R. Svoboda\\
Penn: & W. Frati, K. Lande, A.K. Mann, R. Van Berg\\
U.Texas (Austin) & J.R. Klein\\
\end{tabular}

\vspace{.5in}

\begin{center}
June 16, 2003
\end{center}

\end{titlepage}
\pagenumbering{roman}
\tableofcontents
\vspace{-2ex}
\listoffigures
\vspace{-2ex}
\listoftables
\vspace{-2ex}
\newpage
\pagenumbering{arabic}
\addcontentsline{toc}{section}{Executive Summary}
\addtocontents{toc}{\protect\vspace{.1cm}}
\section*{Executive Summary}

This is a preliminary version of a formal proposal by the 3M collaboration to  construct  a megaton, modular, multipurpose
(3M) neutrino detector with which to perform a program of important experiments in neutrino physics. It will be a program of roughly ten years, during the first part of which the components of
  the complete detector will be constructed in chambers  approximately 7000 ft below the Earth's surface in the Homestake Mine at Lead, South Dakota. During construction and in the second part of this program, the detectors will be used to conduct fundamental experiments on neutrino oscillations that will culminate in the principal experimental goal of the program.
That goal is an experiment to address the profound issue of CP-invariance violation in the lepton sector of elementary particles, an issue that has been the subject of intensive experimental and theoretical study in the quark sector for several decades. This motivation is discussed in section 1 here. The underground chambers and detection apparatus are described at length in later sections of this proposal.
Each component of the final multi-component array is estimated to cost about \$52--54M, including underground chamber costs and contingency.  The yearly expenditure rate for this program should be in the vicinity of \$60 million and  should not deviate greatly from that value.

The principal physics goals of this program also require a moderately intense neutrino beam from an accelerator located at a long distance from the detector array. The distance of BNL from Homestake is 2540 km which provides a span of the values of the physics parameters necessary to carry out the program.  A detailed construction plan 
has been drafted for that purpose and may be found at
 http://nwg.phy.bnl.gov/

The research proposal described here is for more than a single experiment. It involves the construction of an experimental facility that would form the centerpiece of a new deep underground science and engineering laboratory.
 The series of experiments to be carried out in  that facility 
 utilizing neutrino beams far from their accelerator origin will make possible  studies of fundamental features in the lepton sector of elementary particles. The laboratory and the experimental apparatus, apart from  the accelerator neutrino beams, form a plan  whose ambition and cost are beyond that of most of the experimental proposals submitted to the NSF, but all aspects of the plan---including that for the accelerator-generated neutrino beams which will be submitted to the DOE---will survive severe scrutiny.

There are other measurements and experimental searches of importance that can be carried out with the 3M detector independently of and at the same time as the neutrino oscillation and CP-invariance violation measurements are in process. These do not require the accelerator-generated neutrino beam. They are discussed here in the Introduction, section~\ref{intro} as follows: section~\ref{protondecay} Proton Decay,\ 
section~\ref{UHEneut} UHE Neutrinos, \ section~\ref{Supernovae} Supernovae Neutrinos and section~\ref{CP} CP-Invariance Violation. 

The detector for this program is an array of ten individual imaging Cherenkov counters with ultimate total mass twenty times larger than the largest existing detector of its kind. The parameters of some existing and proposed  cherenkov counters are shown in the table.

\newpage

\begin{center}
Executive Summary table:
Comparison of water Cherenkov detector parameters.\\
\begin{tabular}{|l|c|c|c|c|c|c|}
 \hline \hline
{\small Parameters} & {\small Kamiokande} & {\small IMB-3} & {\small Super} & {\small SNO} & {\small UNO} & {\small 3M}\\ 
	& {\small Japan} 	& {\small USA}	& {\small Kamiokande} &	{\small Canada}
& {\small (proposed)} & {\small (proposed)}\\
& {\small II} & & {\small Japan} & & & {\small ten unit} \\
& & & & &  & {\small array} \\ \hline \hline
{\small Total mass} & {\small 4.5 kT} & {\small 8 kT} & {\small50 kT} & {\small8 kT} & {\small 650 kT} & 1,000 kT\\ \hline
{\small Fiducial mass} & & & & & &\\
{\ \ \ \small proton decay} & {\small 1.0 kT} & {\small 3.3 kT} & {\small 22 kT} & {\small 2 kT} & {\small 440 kT} & {\small 800 kT}\\
{\ \ \ \small solar} & {\small 0.7 kT} & - &  {\small 22 kT} & {\small 1 kT} & {\small 440 kT} & 800kT \\
{\ \ \ \small supernova} & {\small2.1 kT} & {\small 6.8 kT} & {\small 32 kT} & {\small 3 kT} & {\small 580 kT} & {\small 800 kT} \\ \hline
{\small Effective} & {\small 20\%} & {\small 4\%} & {\small 40\%} & {\small 60\%} & {\small 1/3 10\%} & {\small 14\%}\\
{\small Photocathode} & & & & & {\small 2/3 40\%} &\\
{\small coverage} & & & & & &\\ \hline
{\small Total size }& {\small $16m \times$} & {\small $22 \times 17 \times$} & {\small $41m \times 
$} & {\small $30 \times $} & {\small $60 \times 60 $} &
{\small 1 unit} \\
& {\small $19 m \phi$} & {\small $ 18m^3$ } & {\small $ 39 m \phi$} & {\small
$22 m \phi$} & {\small $180 m^3$} & {\small $50 \times 50 m \phi$}\\
& & & & & & {\small 10 units}\\
& & & & & & {\small See Fig.~\ref{f:array}}\\
 \hline \hline
\end{tabular}
\end{center}

\newpage

\section{Introduction}
\label{intro}

\noindent The characteristics of the Homestake Underground Laboratory, very deep
laboratory
chambers and very strong rock, open  the possibility of a new generation
of
extremely large, sensitive detectors that can explore hitherto unexamined 
or poorly understood areas of
neutrino physics
and of other fundamental physics interactions. 

Much of neutrino physics to be explored in the Homestake Laboratory is now in the common domain. It has been studied in a preliminary way with smaller, less sensitive detectors and then in more massive, better resolution detectors. 
This was the case, for example, of the atmospheric neutrinos in which vacuum oscillations of muon type neutrinos to (presumably) tauon type neutrinos were first observed and identified in the Kamiokande-II and IMB detectors and subsequently with larger data samples in the Superkamiokande detector. The potential value of studying other neutrino physics has been recognized, but limited by the mass available for detectors, the size of the cavities and shielding to house them, the need for special sensitivity to observe interactions with low neutrino cross sections, and the general expense and difficulty involved in constructing a neutrino detector often not connected directly with a particle accelerator complex. Moreover, most neutrino detectors serve more than one physics purpose; they address a number of physics questions that are distinguished by the energy and type of neutrino incident on them and by the energy and particle types of the products of the neutrino interactions that occur in them. Consequently, in this proposal we stress the design, construction, and cost of a general purpose, state of the art neutrino detector which we label Megaton Modular Multi-purpose
(3M) Neutrino Detector to convey its properties. We illustrate and briefly discuss some of the physics measurements and searches that might be done with the 3M detector but in the main we take for granted that a detector with the 3M properties can be turned to most of the neutrino physics  at present of interest and likely to appear on the horizon in the future.

In particular, the prospect of using  the 3M detector in conjunction with accelerator-produced neutrino beams to carry out extra long baseline experiments has great promise and will be discussed at length here.

Among the recognized scientific goals of such a detector are:
\vspace{3ex}

\begin{enumerate}
\item[] a.  Extend the investigation of neutrino flavor transitions of
atmospheric-type
neutrinos beyond the present explored range  by an extra-long baseline neutrino oscillation experiment employing an intense accelerator neutrino beam and part or all of the 3M array. With this arrangement, it will be possible to demonstrate conclusively that neutrino oscillations are indeed the correct interpretation of the observed anomalous $\nu_{\mu}/\nu_{e}$ ratio in atmospheric neutrinos and in solar neutrinos, and to reach lower, more precise values of the parameters $\Delta m^2$ than heretofore. Moreover, it will be possible  to search for a violation of CP invariance in the neutrino sector. These measurements will be major thrusts of the 3M detector in the next decade; they will be fully described in what follows.

\item[] b.  Search for proton decays one or two orders of magnitude in lifetime
beyond the
present experimental limits, with improved coverage of a variety of possible decay modes. A null result at $10^{35}$ years would  likely  lead to the conclusion that proton decay is not subject to direct detection by present methods.

\vspace{2ex}

\item[] c.  Search for localized or diffuse astrophysical sources of ultra high
energy (UHE)
neutrinos. Such neutrinos have been discussed quantitatively for more than a decade because of the anticipated importance of their possible sources. As yet none have been observed.

\vspace{2ex}

\vspace{2ex}

\item[] d.  Detect neutrino bursts from a supernova and, especially, detect the neutrinos $\nu_e$ (as well as $\bar{\nu}_e$) from the electron capture reaction at the formation of the protoneutron star.

\end{enumerate}

After half a century, the deeper questions addressed by elementary particle
 physicists---although
broader and subtler---are no less difficult to answer, no less directed at
fundamental
issues than the questions they asked when the science began.  One fundamental question concerns the
intrinsic nature
of the neutrino, made even more interesting by the recent observation that
neutrinos of
different type appear to possess small but different  masses and are capable of
spontaneously making a transition from one
type to
another. Another,
because its ramifications are so extensive, concerns the stability of the
proton
against spontaneous decay.

Apart from the profound physics that these questions have in common, they
are also
related by the practical need to extend experimental methods beyond their
present
boundaries in order to address them. This means, on the one hand, extending
the
present lower limit on the proton lifetime beyond its approximate value of
$10^{33}$
years and, on the other hand, searching for neutrino behavior in phenomena
where new
neutrino properties are exhibited, for example, in neutrino oscillation phenomena at extreme values of the parameters, and at neutrino energies
different from  any so far observed or at distances far longer than those at which detailed precision measurements have been made. Modern techniques indicate that a common detector
method
can be employed in both studies provided that the detector can be adequately
shielded
against extraneous phenomena, i.e., cosmic rays and their products. This has
been
demonstrated over the past two decades during which the lower limit above on the proton
lifetime
was acquired and the science of neutrino astrophysics successfully launched.

To go further means appreciably larger, more sophisticated detectors and---particularly
important---significantly improved protection of them against cosmic rays.
This latter factor can be
attained by embedding the detectors deeper in the Earth than heretofore or
deep under
ice, say at the South Pole, or deep in the sea.  A new
underground
laboratory with the features likely to be found in the Homestake
Laboratory opens that possibility on land. In addition,
the large water Cerenkov detectors at Lake Erie, Sudbury and Kamioka,
(Kamiokande~II and
Superkamiokande) have beautifully demonstrated the capability of such
detectors to study
neutrino emission by the Sun, to investigate the time of flight dependence
of neutrino
flavor transitions, to search for baryon number violation via the decay of
the proton and
to observe neutrinos from supernova. The largest of these detectors,
Superkamiokande (SK),
has a total water volume of ${\rm 50,000 m^{3}}$ and is at a depth of
approximately 2700 mwe.

We visualize an array of ten  detectors for the Homestake
Laboratory, each of 100 kilotons (50m diam $\times$ 50m  high in
volume), with a total detector mass of the order of a megaton that would serve as a target to provide high statistics observations of reactions produced by intense sources of neutrinos for extra-long baseline neutrino oscillation experiments and for searches for CP violation in the neutrino sector. The fiducial volume of each of the 100 kiloton detectors will be about 80 kilotons. The array will
ultimately extend the range of proton decay lifetime searches by two
orders of magnitude from the present value and simultaneously serve as
a hodoscope for ultrahigh energy (UHE) neutrino detection.
The array would also provide a high statistics observation of the principal neutrino emission from a supernova and, for example, one of the array filled  with a low threshold material for neutrino interactions, e.g., $^{37}Cl$ in the form of NaCl, would make possible observation of the $\nu_e$ from $e^- + p \rightarrow n + \nu_e$ at the formation of the protoneutron star.
This array of detectors, shown schematically in Fig.~\ref{f:array}, allows for all of its
modules  to be used in concert, i.e., as a megaton detector for most experiments, e.g., proton decay, and UHE neutrino searches, supernova observations, and CP violation search, while an individual member or a small number of members of the array is devoted to a different but related purpose, e.g., study of neutrino oscillation parameters or  detection of the $\nu_e$ as well as the $\bar{\nu}_e$ from a supernova.

\par\jeanfig{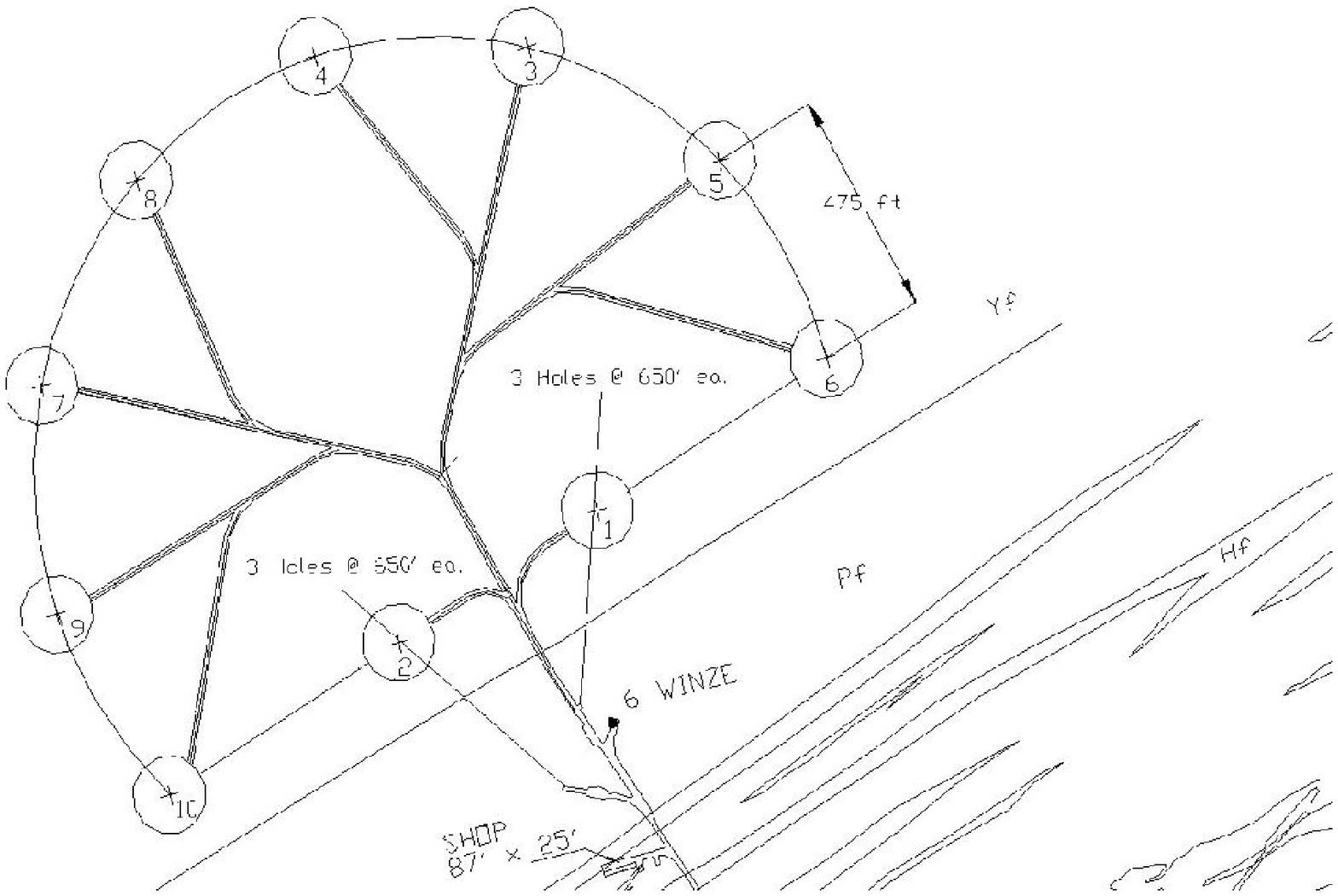}{4}{View of the 3M detector array showing the top access (6950 ft. level of the Homestake Mine.) The modules are right
 circular cylinders 50 m in diameter and 50 m high.}

\par\jeanfig{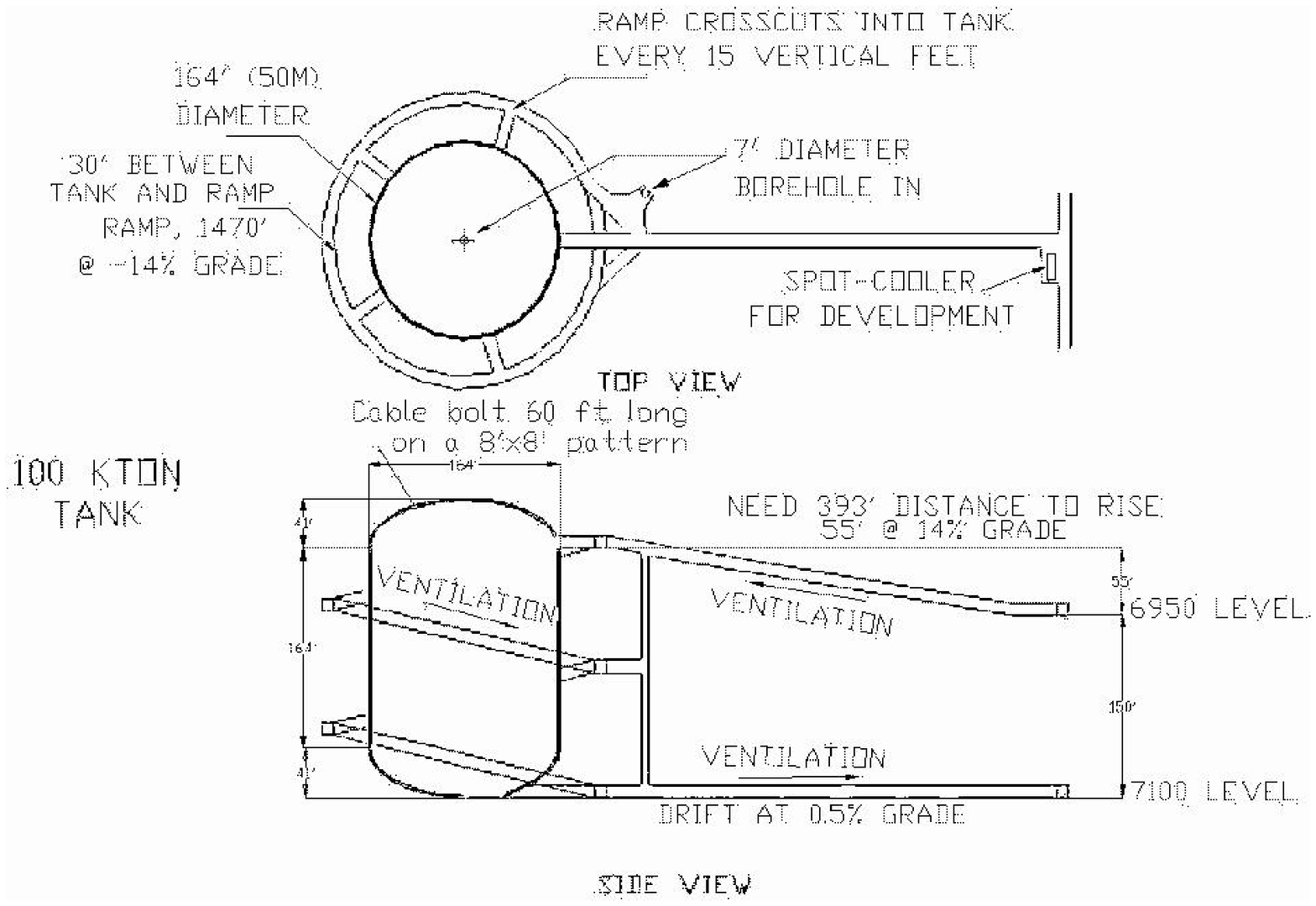}{3.5}{Elevation view of a single detector module of the 3M detector showing the access tunnels, sizes and levels.}
 Locating
the array at a depth of, say, 7,000 ft. (6200 mwe) would result in a cosmic ray
muon flux background per module that is more than two orders of
magnitude less than that at the SK detector and a total muon
background for the entire megaton array that is an order of magnitude
lower than the background in the 50 kton SK detector.
Considering the successes of SK and SNO (at 2700 and 6210 mwe, respectively) in coping with
the cosmic ray background at their respective sites, 
 the water cerenkov detectors in the Homestake laboratory
might be operated without any, or at least a minimal, loss of useful
volume occupied by all-enclosing veto-counters.  The cosmic ray muon background  at 6200 mwe will be as shown
in Fig.~\ref{f:muons}.
\par\jeanfig{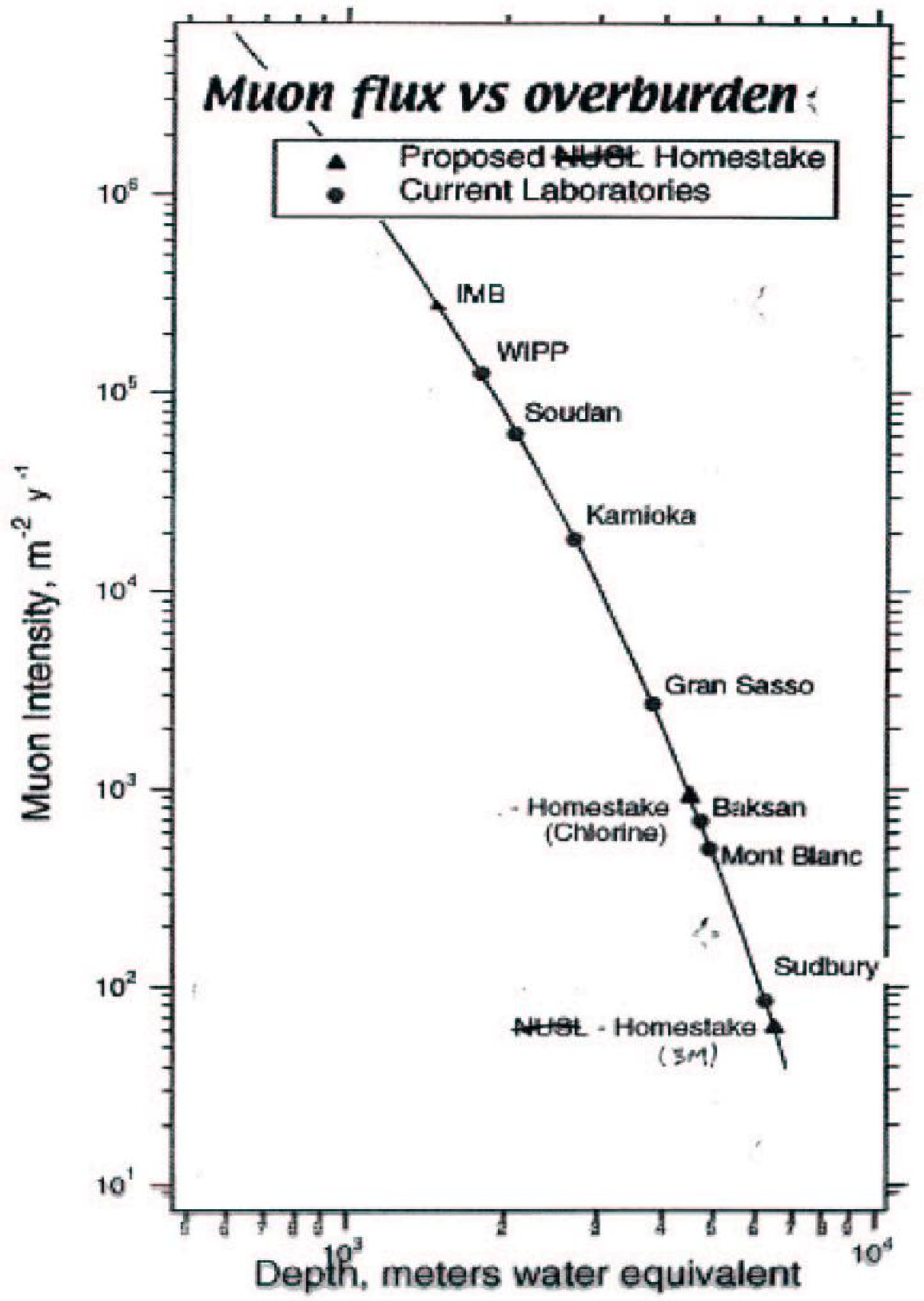}{3.5}{Cosmic ray intensity as a function of underground depth. At Homestake, the representative measured fluxes are 4 muons per $m^2$ per day at 4850 ft., 0.4 muons per $m^2$ per day at
 6950 ft., and 0.13 muons per $m^2$ per day at 8000 ft.}

When, in addition, the detector is used with a pulsed neutrino beam from an accelerator, the cosmic ray muon background is negligible.

\subsection{Proton Decay}
\label{protondecay}

As indicated above, the mass of a single cavity or chamber of the
detector array that we have in mind is about 100 kilotons of ${\rm H_{2}O}$. 
We
think that excavating and completing one chamber of specified volume, properly rock-bolted, with adequate ventilation and lighting, and encased walls-- as soon as possible after
formation of the underground laboratory is necessary to gain early experience in the construction of such large chambers. The first detector component may soon thereafter be adequate to  measure the energies of the
charged products of  neutrino interactions and of possible products of proton decay.

More
completed detector units will be added to the proton-decay
complex.  A
conservative estimate suggests that a complex of at least five chambers
can be excavated  and accept Cerenkov counters within a
period of    seven years for less than \$300 million (see below for details).  They will
have been equipped for particle detection  as  discussed below; some of them at the same time as excavation of the later ones is proceeding, so that data can be taken continuously after the first module is completed.

Building on the previous work on proton-decay and  atmospheric neutrino interactions of the earlier water
Cerenkov detectors at Kamioka, viz., Kamiokande \cite{hirata},
Kamiokande-II \cite{KII}, Superkamiokande (SK) \cite{SK}, and at
 the Fairport Salt Mine \cite{IMB} under Lake Erie, few new problems in
analyzing  data from the Homestake complex should arise, and summing the
output of the Homestake detectors should yield an ultimate lower limit on
restricted decay
modes of proton-decay about two  orders of magnitude higher
than the present SK limit ($10^{33}$yr).

\subsection{UHE Neutrinos}
\label{UHEneut}

There are several quantitative predictions of the possible flux of UHE
neutrinos from various astrophysical and cosmological remnant
sources. The predictions cover a wide range of neutrino energies and
fluxes from a few TeV for the known atmospheric neutrinos to roughly
$10^9$ TeV for neutrinos suggested by speculations on topological
defects; some of the predictions are shown in Fig.~\ref{f:UHE} \cite{nellen}.
\par\jeanfig{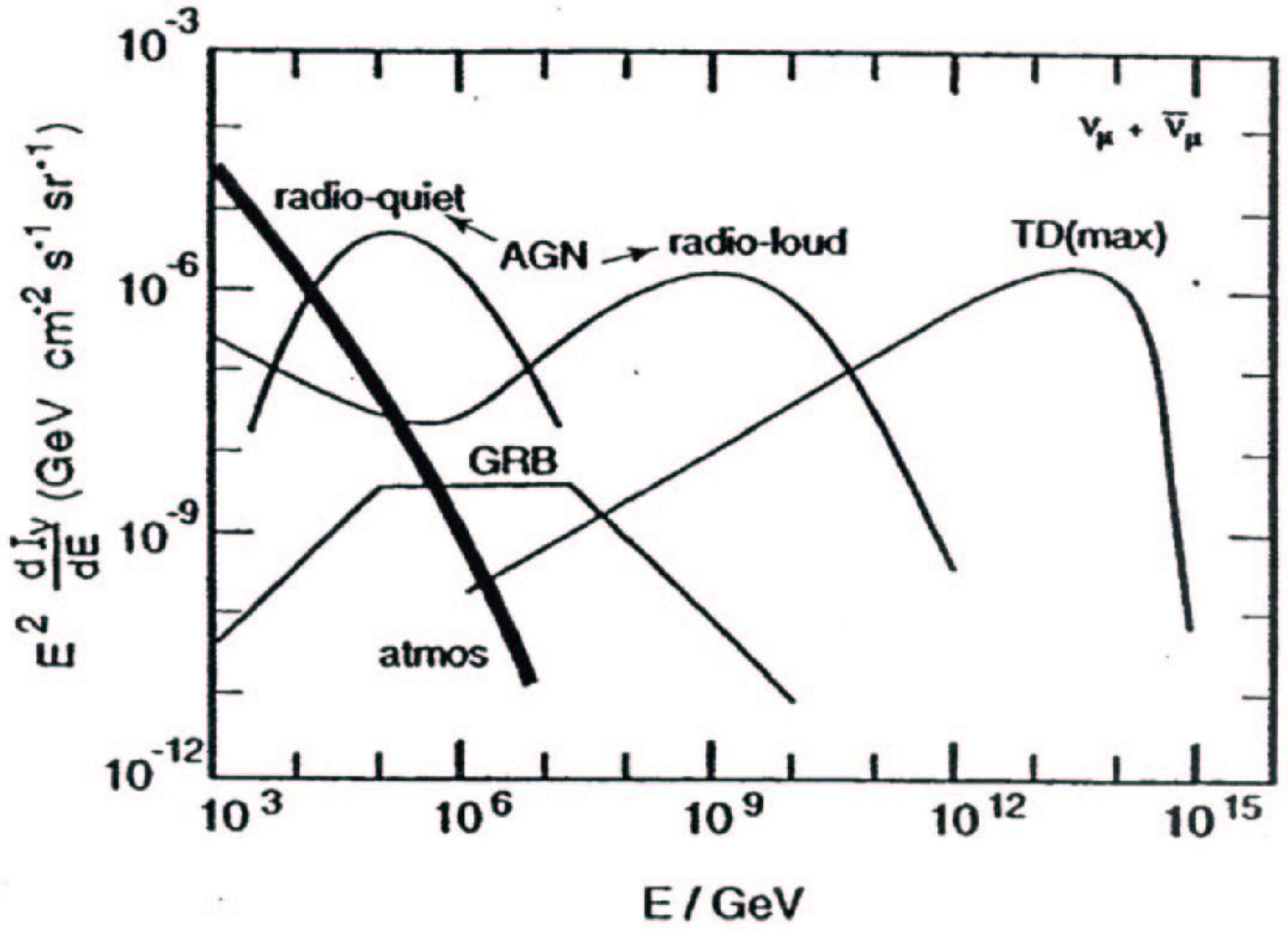}{3.5}{Calculated UHE neutrino (plus antineutrino) fluxes at the Earth's surface weighted by $E^2$. See reference \cite{hirata}.}
A clear positive result in the search for UHE neutrinos would open new
unexplored areas in physics, astronomy and cosmology. The modular (3M)
detector is well-suited for such a search. Roughly, we
expect a large UHE muon-background-free region in the angular interval
$70^0 < \theta < 120^0$, where $\theta$ is the zenith angle defined by the
normal to the plane of the detectors in Fig.~\ref{f:array}. A UHE neutrino-induced
muon signal, for example, will traverse between one and three modules
depending on their final spatial configuration. The modules will act as a sampling
detector to roughly determine  muon direction and muon energy within an order
of magnitude or better. We expect the detector in Fig.~\ref{f:array}, depending
on final dimensions, to subtend a solid angle $\Delta \Omega$ relative to the solid
angle subtended by the canonical ${\rm km^3}$ detector $\Delta \Omega ({\rm km}^3) $ often referred to
in \cite{kleinandmann} 
of $0.01 < \Delta \Omega/ \Delta \Omega ({\rm km}^3) <
0.10$. That relative solid angle interval would be satisfactory for an
initial survey in which the goal would be to observe a small number of
events to show that UHE neutrinos exist. A positive signal from the
proposed detector or any other would be impetus to expand the array in
Fig.~\ref{f:array} further.

 It is realistic to consider
the prospect of observing  low intensity fluxes such as shown in Fig.~\ref{f:UHE} because the
reaction cross section for neutrino plus nucleon rises steeply with
increasing neutrino energy for both neutrinos and antineutrinos. For
example, the cross section rises by five decades as the neutrino
energy increases by seven decades despite the production of real
intermediate vector bosons. This is given in more detail in Fig.~\ref{f:crosssection} \cite{kleinandmann}.
\par\jeanfig{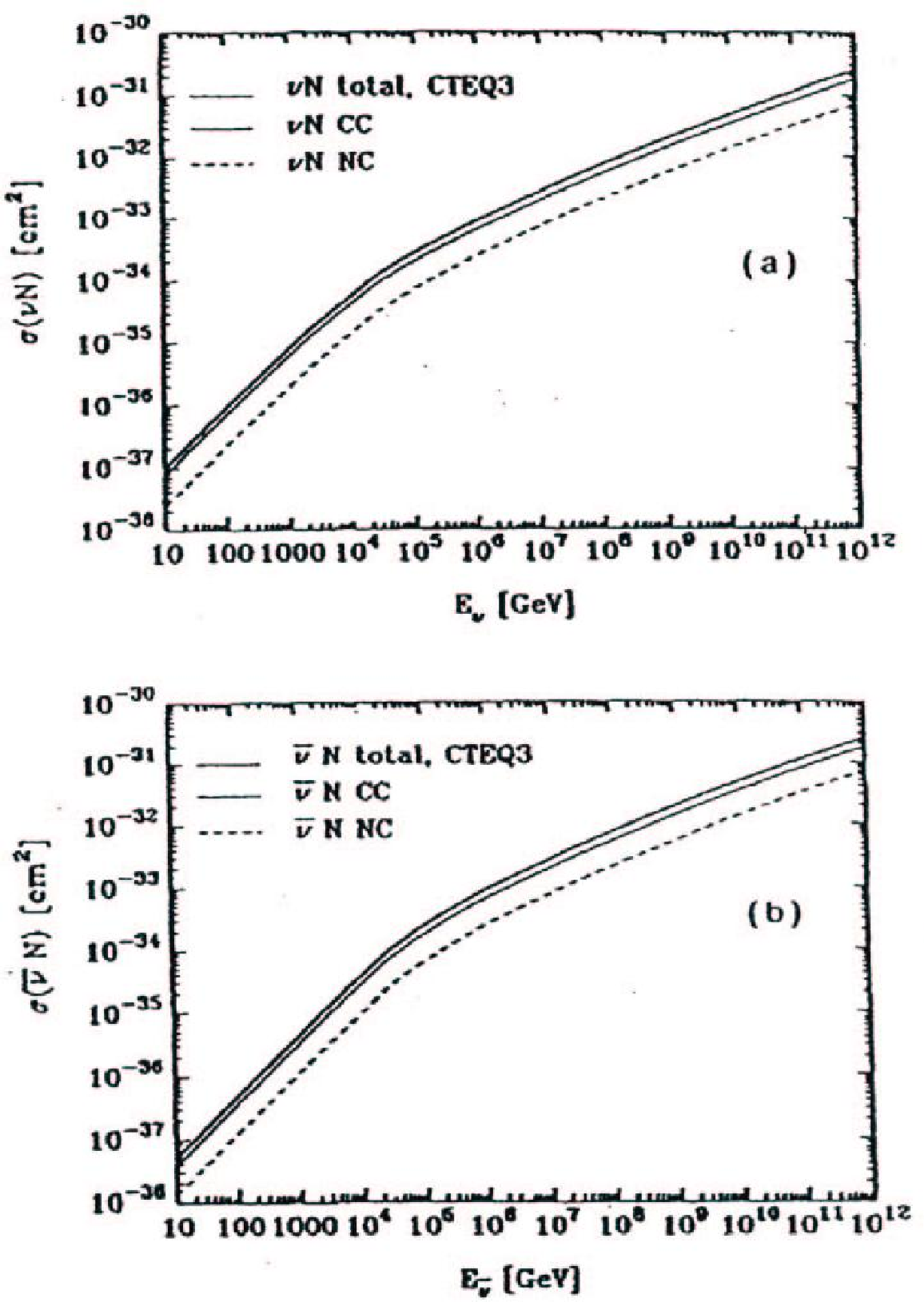}{3.5}{Cross sections for (a) $\nu N$ interactions at high energies: dotted line, $\sigma (\nu N \rightarrow \nu +$ anything); thin line, $\sigma (\nu N \rightarrow \mu^- $ + anything); thick line, total charged current plus neutral current cross section. (b) for $\bar{\nu} N$ interactions. See reference \cite{KII}}
An important consequence of the rapid increase of neutrino interaction
cross sections is that the Earth becomes a significant absorber of
neutrinos above roughly $10^4$ GeV and is essentially opaque to
neutrinos with energy above $10^7$ GeV. The neutrino survival
probability as a function of $\cos \theta_Z$ for those neutrino
energies
is plotted in Fig.~\ref{f:survival}; $\theta_Z$ is the zenith angle at the
detector \cite{AKM}.
\par\jeanfig{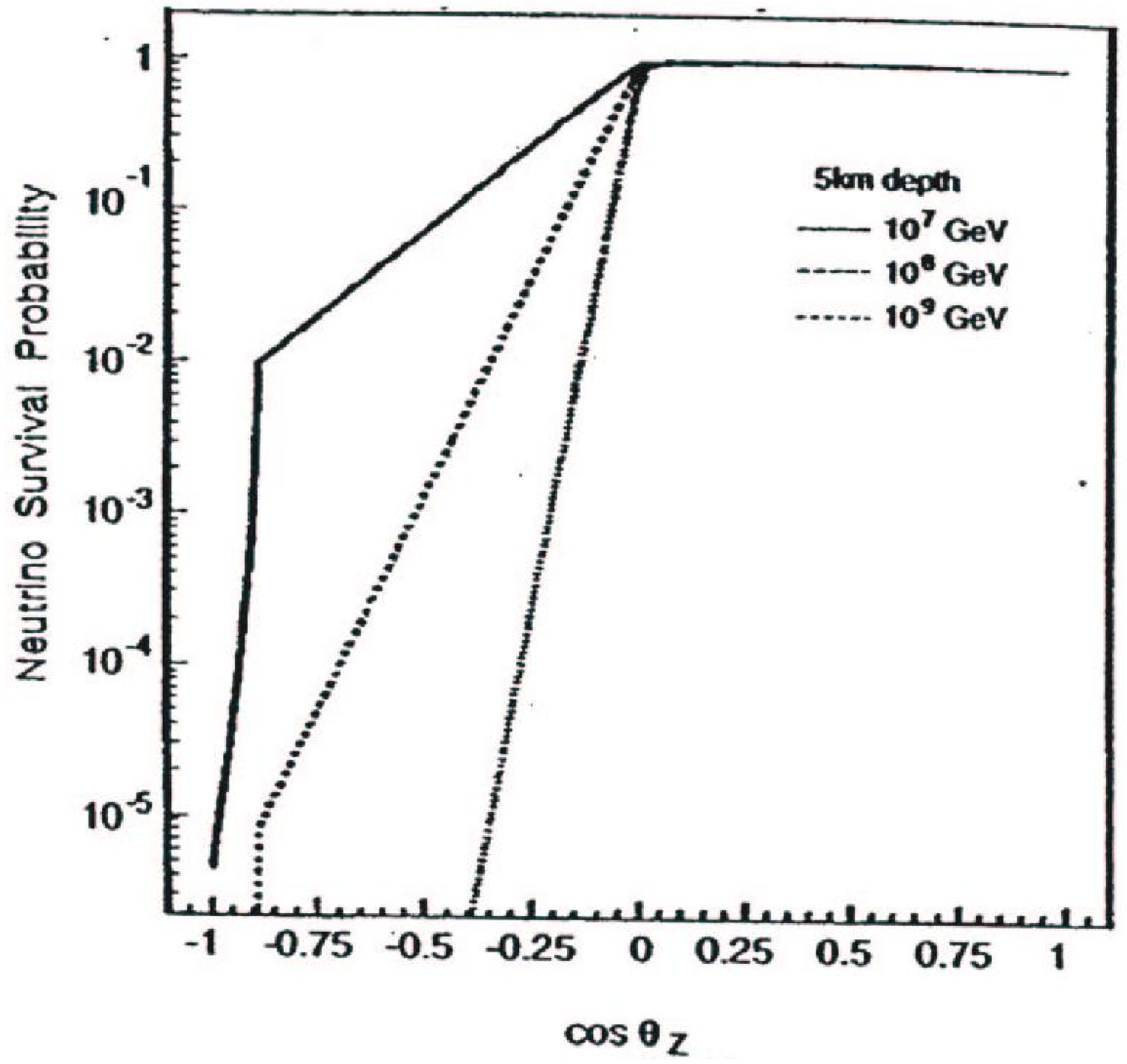}{3.5}{Survival probability of UHE neutrinos from space as a function of $\cos \Theta_Z$, the zenith angle at a detector a few kilometers below the Earth's surface; $\cos \Theta_Z = +1$ corresponds to the zenith direction. The sharp break in slope for the lower energy neutrinos at $\cos \Theta_z = -0.9$ occurs when they emerge from the shadow of the Earth's core. See reference \cite{SK}}.

 In obtaining the curves in Fig.~\ref{f:survival}, the Earth is modeled as
a high density  ($ {\rm 15 gm/cm^3}$) core and a low density mantle,
which accounts for the sharp break near large negative values of $\cos
\theta_Z$ at the lower energies. This effect requires serious
consideration in 
formulating plans for UHE neutrino searches, especially if the actual
fluxes are much lower than suggested by Fig.~\ref{f:UHE}. 

A less dramatic, but nevertheless highly interesting, measurement that also becomes possible with the array and the increase of neutrino cross sections is the mapping of the density of the Earth's interior as described briefly in Fig.~\ref{f:density} \cite{AKM}.
\par\jeanfig{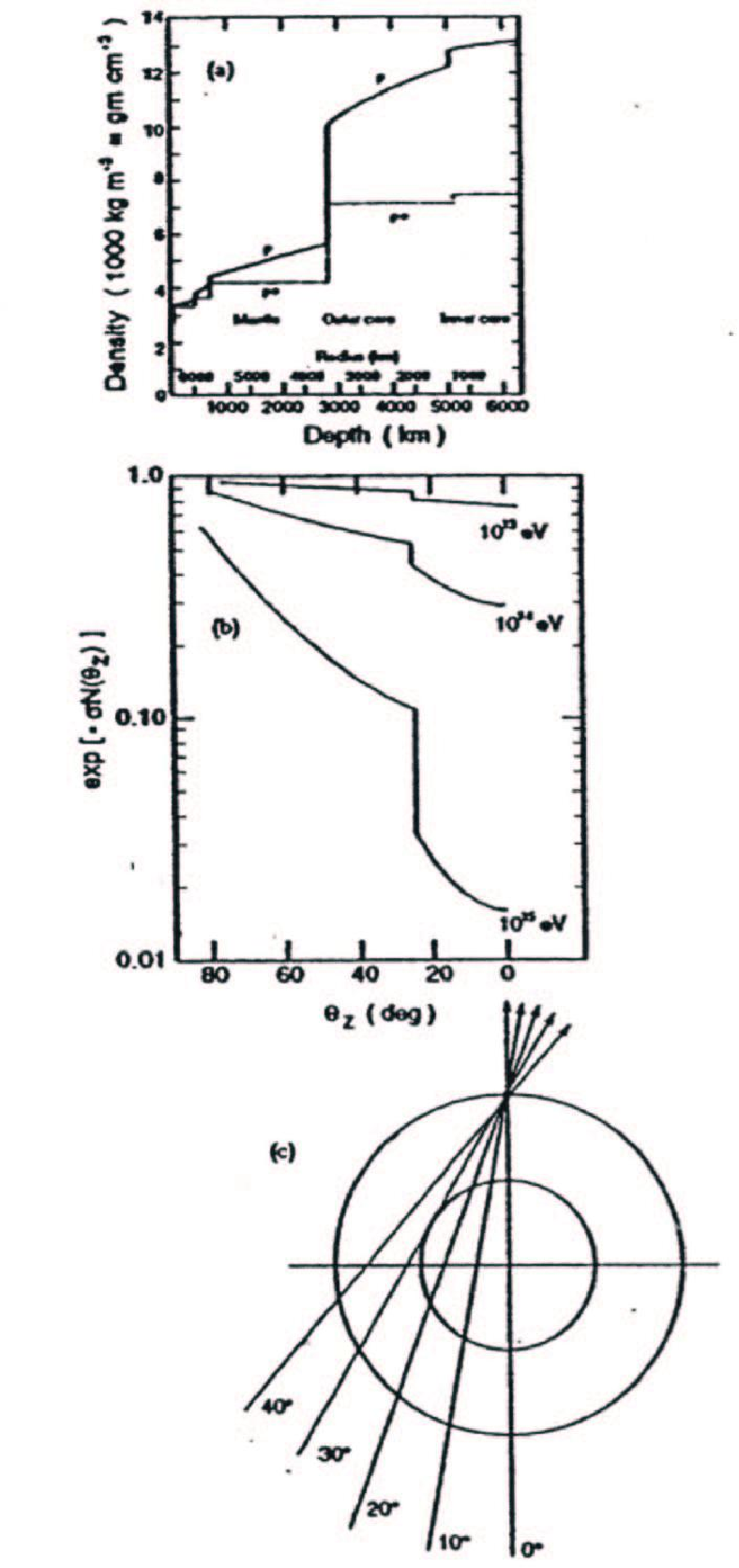}{7.0}{(a) A model of the density profile of the earth (from F.D. Stacey), (b) Crude simulation of the central region, 3,000 to 6,000 km and outer region, 0 to 3,000 km, showing idealized neutrino paths. (c) Attenuation of the neutrino flux from the lower hemisphere in (b), i.e., $\exp[-\sigma N (\theta_Z)]$ \underline{vs} $\theta_z$ assuming density of the central region to be $15 gm/cm^3$, and of the outer region to be $4 gm/cm^3$. Note that for $\theta_z > 25^0$, which corresponds to the edge of the central region, the attenuation is assumed to be determined solely by the path length at a fixed density.
See reference \cite{IMB}}.

\subsection{Supernovae Neutrinos}
\label{Supernovae}

Since the first observations of neutrinos from the type II supernova \cite{sn1987}, SN 1987A, the intent of neutrino astrophysicists is to be prepared to study the neutrino emission from a type II supernova in our own or another nearby galaxy if and when the occasion arises.  One of the detectors in the array in Fig.~\ref{f:array} would provide a good statistical sample of such neutrinos for analysis; however, utilization
of another member of the array for a different but correlated measurement might be particularly rewarding if the second detector module is properly equipped.

The current model of a type II supernova core collapse involves a step in which the protoneutron star is formed through complete dissociation of the components of the nuclei in the collapsed iron core followed by a newly freed proton  capturing an electron;  the reaction giving rise to a neutron and  $\nu_e$. 
The resulting burst of $\nu_e$ was not observed during SN 1987A because it is expected to be short-lived ($\leq 1 sec$), and to carry only about ten percent of the total neutrino energy emitted by the supernova in the subsequent $\nu_e$,
$\bar{\nu}_e$, etc. pairs. Measurement of the short $\nu_e$ pulse would mark the instant from which to measure all later times of interest in the event.

However, in a water Cerenkov detector, the low energy  $\bar{\nu}_e$
(antineutrinos) from a supernova are much easier to detect than the $\nu_e$ because the two protons in the water molecule are essentially free to participate in the reaction $\bar{\nu}_e + p \rightarrow e^- + n$, while the neutrons which are required to satisfy separate lepton number and charge conservation in the particle reaction $\nu_e + n \rightarrow e^- +p$ are tightly bound in the oxygen nucleus and accordingly demand higher energy neutrinos to make the latter particle reaction go. The $\bar{\nu}_e$ from SN 1987A were in fact the ones detected.

If, however, a second detector module contained chlorine in the form, say, of $NaCl$  the reaction $\nu_e + ^{37} Cl \rightarrow e^- + ^{37}Ar$, the same as observed in the pioneer solar neutrino experiment \cite{davis}, would occur in it
 with a neutrino energy threshold of 0.814 MeV; and electrons with kinetic energy above roughly 7 MeV from that reaction would be detected as would their emission times and energies.  A comparison of the energy and time spectra of the $\nu_e$ and $\bar{\nu}_e$ induced electron spectra from the two reactions would describe the direct neutrino emission from the supernova completely with at least adequate statistical samples.
Diversion of one of the 100 kiloton modules for this purpose would not render it unuseful for the other experiments described here.

\subsection{CP-Invariance Violation}
\label{CP}

The physics  reach of the 3M program in the study of CP-violation
is extensive and needs to be presented in quantitative detail to be
 appreciated. A compact but complete presentation of that material is soon to be published in Physical Review D and is available at arXiv: hep-ph/0303081 v2
 16 May 2003. We reproduce here only parts of the Introduction and Conclusions of that report.

The first violation of CP-invariance was observed in the $K^0-\bar{K}^0$ system about forty years ago \cite{fitch}. Until that experiment, it was thought that the product of those discrete symmetry factors was invariant and preserved the symmetry between a fundamental neutral particle and its antiparticle. Recently, a CP violation has been observed in a second quark system, the $B^0-\bar{B}^)$ system \cite{bquark}. Following that, interest is raised in the question of the presence of CP-invariance violation in the neutrino sector.

The observation of neutrino oscillations opens the neutrino sector as a promising region in which a CP asymmetry between neutrino and antineutrino might be exhibited. This would fuel further speculation that a primordial universal asymmetry in the early universe accounts for the matter-antimatter asymmetry we observe now. The suggestion that a long baseline neutrino oscillation experimental program would be a fertile region to explore for that purpose has been studied intensively \cite{marciano} in the last few years as a result of the experimental data on neutrino oscillations but a conclusive demonstration of neutrino oscillations and a search for CP-invariance violation await a more ambitious experimental program such as is proposed here.

\vspace{1ex}

\begin{center}
{\it Introduction}\\
{\it from `` Very Long Baseline Neutrino Oscillation Experiments for Precise 
Measurements of Mixing Parameters and  CP Violating Effects'', to be published
in Physical Review D, arXiv: hep-ph/0303081 v2 16May 2003.}
\end{center}

\vspace{1ex}

{\it Measurements of solar and atmospheric neutrinos have provided strong
evidence for non-zero neutrino masses and mixing
[1-3].  Atmospheric results have been further
strengthened by the K2K collaboration's accelerator based results[4].
 The Solar neutrino results have been confirmed by the
KamLAND collaboration in a reactor based experiment that has shown
that the large mixing angle (LMA) solution is most likely the correct
one [5].  Interpretation of the experimental results is
based on oscillations of one neutrino flavor state, $\nu_e, \nu_{\mu}$
or $\nu_{\tau}$, into the others, and described quantum mechanically
in terms of  neutrino mass eigenstates, $\nu_1,
\nu_2$ and $\nu_3$. The mass squared differences  involved in
the transitions are measured to be approximately $\Delta m^2_{21}
\equiv  m {(\nu_2)}^2 - m {(\nu_1)}^2  = (5 -20) \times 10^{-5} {\rm
eV}^2$ for the solar neutrinos and $\Delta m^2_{32} \equiv
m{(\nu_3)}^2 - m{(\nu_2)}^2 = \pm(1.6 - 4.0) \times 10^{-3} {\rm eV}^2$
for the atmospheric neutrinos, with large mixing strengths, $\sin^2 2
\theta_{12}\approx 0.86$ and $\sin^2 2 \theta_{23}\approx 1.0$ in both
cases. 
In this paper, we show that
the currently favored parameters open the possibility for an
accelerator based very long baseline (BVLB) experiment that can explore both
solar and atmospheric oscillation parameters in a single experiment,
complete the measurement of the mixing parameters with good precision, and search for new
physics.}

\vspace{1ex}

\begin{center}
{\it Conclusions}
\end{center}

\vspace{1ex}

{\it We have simulated and analyzed a feasible, very long baseline  neutrino
oscillation experiment consisting of a low energy, wide band neutrino
beam produced by 1 MW of 28 GeV protons incident on a carbon target
with magnetic horn focusing of pions and kaons, and a 500 kT detector
at a distance of $>$ 2500 km from the neutrino source. The BVLB neutrino
beam with a total intensity of about $4.7\times 10^{-5} \nu/m^2/POT$
at a distance of 1 km from the target could be provided by an upgrade
to the BNL-AGS [17].

 The single BVLB experiment could produce measurements of all parameters
 in the neutrino mixing matrix through observation of the
 disappearance channel, $\nu_{\mu} \rightarrow \nu_{\mu}$ and the
 appearance channel, $\nu_{\mu} \rightarrow \nu_e$.  
The experiment is also sensitive to the mass
 ordering of neutrinos using the observation of the matter effect in
 the appearance channel through the currently unknown parameter
 $\sin^2 2 \theta_{13}$. Nevertheless,  the experiment is intended
 primarily to measure the strength of CP~invariance violation in the
 neutrino sector and will provide a measurement of the CP~phase,
 $\delta_{CP}$ or alternatively the CP~violating quantity, $J_{CP}$, 
  if the one currently unknown neutrino oscillation
 mixing parameter $\sin ^2 2 \theta_{13} \geq 0.01$, a value about 15
 times lower than the present experimental upper limit. We
 point out that for a given resolution on $\delta_{CP}$ the number of
 neutrino events needed which determines the detector size and beam
 intensity, is approximately independent of the baseline as well as
 the value of $\sin ^2 2 \theta_{13}$ as long as the electron signal
 is not background dominated. Therefore,
the concept of very long baseline ($\ge
 2000~km$) is  attractive because it provides access to much
 richer physics phenomena as well as reduces the need to understand
 systematic errors on the flux normalization and background
 determination. A shorter baseline ($<1500 ~km$) will obviously limit
 the reach of the experiment for $\Delta m^2_{21}\times \sin 2
 \theta_{12}$ in the appearance mode as well as reduce the 
resolution on the CP parameter. On the other hand, a much longer
 baseline ($>4000 ~km$) will result in a matter effect that is large
 enough to dominate the spectrum and make the extraction of the
 CP~effect more difficult. The larger distance will also make it
 necessary to make the neutrino beam  directed
 at a higher angle into the Earth, a technical challenge that may not be  
necessary given the current values of neutrino parameters.
Lastly, we have also shown that most of this rich physics
 program including the search for CP~effects can be carried out by
 neutrino running alone. Once CP~effects are established with neutrino
 data, anti-neutrino data could be obtained for more precision on the
 parameters or search for new physics. The shape of the 
disappearance signal over multiple oscillations
from   neutrino running alone (as well as in combination with 
anti-neutrino running) can be used to contrain effects of new physics
from: sterile neutrino mixing, extra dimensions, exotic interactions
with matter, etc.

 It has not escaped our notice that the large detector size ($\ge$500
 kT), mandated by the above described neutrino program, naturally
 lends itself to the important physics of proton decay, supernova 
detection, etc. Although some
 experimental requirements such as shielding from cosmic rays,
 detector shape, or photo-sensor coverage, may differ, it is clear
 that they can be resolved so that this large detector can be employed
 to address these two frontier physics issues.}

{\it References}
\begin{enumerate}
\item[] 1. {\it PDG, Phys. Rev. {\bf D66}, 010001 (2002), p. 281.}

\item[] 2. {\it Y.~Fukuda et al., Phys. Rev. Lett. {\bf 81}, 1562 (1998);
S. Fukuda et al., Phys. Rev. Lett. {\bf 86} 5656, 2001;  
E.W. Beier, Phys. Lett. {\bf B283}, 446 (1992); 
T. Kajita and Y. Totsuka, Rev. Mod. Phys. {\bf 73}, 85 (2001).}

\item[] 3. {\it  Q. R. Ahmad et al., Phys. Rev. Lett. {\bf 87} 071301 (2001). 
S. Fukuda et al., Phys. Rev. Lett., {\bf 86} 5651 (2001).} 

\item[] 4.
{\it S. H. Ahn et al., Phys. Lett. {\bf B 511} 178 (2001).}

\item[] 5. {\it  K. Eguchi  et. al, Phys. Ref. Lett. {\bf 90},
  021802 (2003). hep-ex/0212021.}

\item[]17.  {\it J. Alessi  et al., AGS Super Neutrino Beam 
Facility, Accelerator and Target system Design, BNL-71228-2003-IR. 
April 15, 2003.  http://nwg.phy.bnl.gov/}
\end{enumerate}

We now turn to the essentials of the 3M detector array.

\section{Summary of Construction Planning, Budget and
 Timetable for the 3M Rooms in the Homestake Mine \label{summ_const}}

Plans for the construction of the chambers for a multiple module megaton Cerenkov
detector at the Homestake Mine have gone through three  evaluation and design stages consisting of rock strength and stability evaluation, chamber design and layout, construction planning, and sequencing and development of budget and timetable. Here is a summary of these steps.

\subsection{Characteristics of Homestake Rock, Stability of Large, Deep Caverns
and Coring Plans for the Deep Laboratory Area}

With the exception of several years during World War II, the Homestake Gold Mine in Lead, South Dakota has been in continuous operation since 1876 \cite{Homestakebook}. It ceased active operation in 2001. During that period, over 40 million ounces of gold have been produced. This production involved the excavation of several hundred million tons of ore from the underground mine as well as over 170 million tons of excavated ore and waste rock from the Open Cut (pit) surface mine.

\begin{figure}
\begin{center}
{\includegraphics[width=6in]{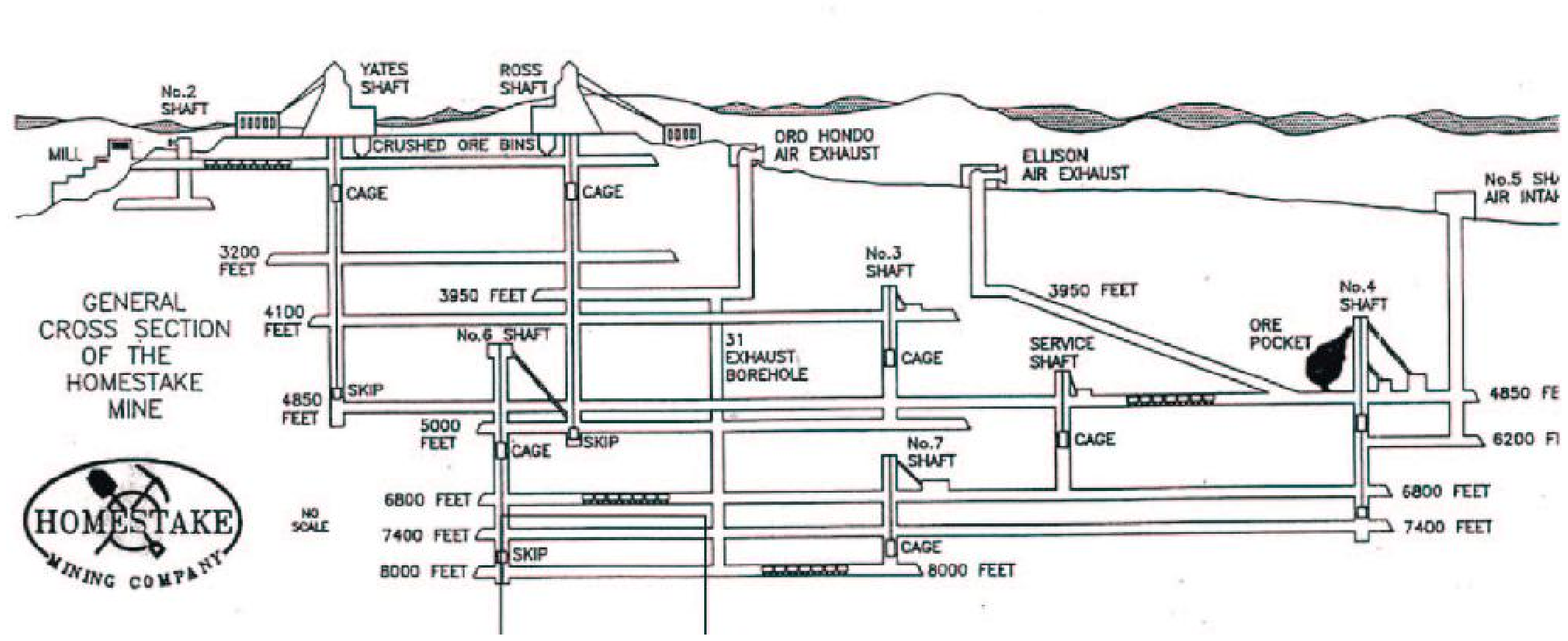}}
\caption{\label{tunnels}Map showing shafts and tunnels at various depths in the Homestake Mine.}
\end{center}
\end{figure}

The Homestake Mine now involves over 160 miles of underground tunnels on numerous levels down to 8000 ft. below the surface, as shown in Fig.~\ref{tunnels}. 
The Homestake Mine geology has been extensively studied (see the U.S. Geological Survey Bulletin 1857-J (1991) and the references cited therein). The strength characteristics of the rock have been thoroughly studied and measured. The advantage of locating detectors deep underground is shown quantitatively
 in Fig.~\ref{f:muons}.
The Homestake Mining Company (HMC) has constructed several large,
 deep underground chambers. Among these are an equipment repair shop at the 7400 ft. level and an air conditioning plant at the 6950 ft. level.
These excavations at depths greater than 7000 ft. provide a strong indication that large excavations at depths of 6500 ft. to 7000 ft. can be constructed and will remain stable for  multi-decade periods. The construction of such
an array of very large chambers also provides a unique opportunity for the rock mechanics and geophysics communities to monitor the behavior of such structures for long periods. This would involve pre-excavation computer modeling of the structure, the installation of strain and stress gauges during construction and the post-excavation monitoring of these instruments. The backfilling of excavations in normal mining precludes long term monitoring and the irregular shapes of ore deposits limits the geometrical range of the excavations that can be studied.

The region of rock being considered for the deep underground 3M detector has not been mined or explored. By extrapolating from above and from the west side, this rock region appears homogeneous, and with few intrusions. Before beginning any excavations, it would be advisable to carry out coring studies of this volume of rock. Cores can be drilled down from tunnels on the 4850 ft. level and sideward and diagonally upward from the 7400 ft. level. These cores would verify the rock strength characteristics and would provide information about the direction of rock grain.

\subsection{Determination of Excavation Stability}

A preliminary 2 dimensional large chamber stability evaluation was carried out in the fall of 2000 by  members of 
the Rock Stability Group at the Spokane Research Laboratory of NIOSH (National Institute of Occupational Safety and Health). This evaluation indicated that stable chambers with dimensions in excess of 50 - 60 meters could be constructed at depths of 7000 ft or more at the Homestake Mine. In the fall of 2001, rock samples from the Yates formation were taken to Spokane and strength and stress analysed.  These measurements provided more specific input for a three dimensional stability analysis of  large excavations as a function of depth in the Yates rock formation in the Homestake Mine. This 3D analysis involved a  finite difference evaluation using the FLAC3D program. These results \cite{initialstability} were compared with the empirical prediction charts of Barton and Grimstad. The conclusions were that 50 meter diameter by 50 meter high chambers could be safely excavated and would be stable for long term occupancy at 2150 meter depth and probably somewhat deeper.

The results of this evaluation were presented in a paper at the Fifth North American Rock Mechanics Symposium in July 2002 \cite{initialstability}.  For the measured Yates formation rock parameters (Tunneling Quality Index), the Grimstad and Barton specifications are 4 - 5 cm of unreinforced shotcrete applied to the rock surface.  We decided to install a 50 cm thick concrete shell on the vertical cylindrical inside of the excavation.  This is equivalent to a two order of magnitude safety factor using the Barton and Grimstad parameters.

The FLAC3D analysis also indicated that in order to isolate the stress fields of one chamber from the adjacent one, these chambers should be separated by at least 3 chamber diameters.  That specification then determined the radius of the circle  on which the chambers will reside.  Finally, the circular layout was adjusted to avoid known inhomogeneities in the rock. See Fig.~\ref{f:array}.  
These conclusions were confirmed by the report of the NSF site selection
committee \cite{siteselection}.

\subsection{Construction of Multiple 100 kiloton Chambers\\
 in the Homestake Mine}
\label{multiple}

Using the results of the stability evaluation, a group of ex-Homestake mining engineers, (Mark Laurenti--former Chief Mine Engineer, Mike Stahl--former Mine
Production Engineer and John Marks--former Chief Ventilation Engineer) designed
an array of ten 100 kiloton water Cerenkov chambers. The criteria used in this design were, a minimum of 50 year safe occupancy of the chambers, independent ventilation and access system for each chamber so that completed chambers can be used for research while additional chambers are under construction, and a structure that will permit a plastic lined water tight and radon reducing structural enclosure.

The design involved a detailed construction plan, a rock reinforcement plan that involved cable and rock bolts and a cylindrical concrete liner, and a coordinated water handling, ventilation and chamber access plan.  The chambers were laid out on the circumference of a 250 meter radius circle that is centered on the Winze 6 shaft. The top of each chamber is connected to the 6950 ft. station of the shaft via a horizontal, radial tunnel. A similar tunnel connects the bottom of each chamber to the 7100 ft. shaft station. Fresh air will be sent to each chamber via the top tunnel and exhaust air removed via the bottom tunnel, thus providing independent air supplies to each chamber. Plan and elevation views of the ten chamber layout are shown in Figs.~\ref{f:array} and ~\ref{f:module}.

A single radial tunnel will access a group of three chambers.  This tunnel is directed toward the central chamber of the group with side branching tunnels to the other two chambers of the group.  The branching locations are staggered to avoid a three tunnel meeting point.  The first step in construction will be to simultaneously drive the main access tunnels at the 6950 and 7100 ft levels.  An equipment maintenance room will be constructed alongside the top access tunnel.  After construction of the three chamber group is completed, this room will be converted into an electronics and water purification facility that serves this three chamber complex.

Two 2.4 meter diameter vertical bore holes will be constructed between the upper and lower tunnels.  One of these bore holes will be outside the chamber and the second in the center of the chamber.  The bore holes serve three purposes.  One, they complete the air flow path between the fresh air input of the upper tunnel and the exhaust of the lower tunnel.  Two, they provide a drop path for excavated rock to the lower tunner for removal.  Three, the bore hole in the center of the chamber provides an expansion region for blasting of rock.  By blasting rock inward from an ever expanding set of concentric circles, we will avoid fracture of the outer rock walls that will serve as the chamber containing walls.

Removing waste rock and associated rock dust via the bottom, air exhaust tunnel, will prevent rock dust from one chamber contaminating the fresh air supply of another chamber. Once construction is completed, the bottom chamber to tunnel connection will be sealed. 

Once the access tunnels and bore holes are completed, construction of the dished dome of the chamber will begin.  The first step is to excavate a 50 meter diameter by 3 meter high section.  80 ft. long vertical cable bolts will be installed into the roof of this excavation as that work proceeds. Once this section is completed, the excavation will be extended upward  to form the dome. As this domed excavation progresses, some of the 80 ft. cable bolts will be cut.  When completed, the dome will have  60 ft. cable bolts in the center and longer cable bolts at larger radii with 80 ft. cable bolts along the outer edge of the dome.  As the stress diagrams in ref. \cite{initialstability} indicate, the regions of greatest stress are along the outer region of the dome, where the cable bolts are the longest.

Once the central chamber access tunnels and boreholes are completed, work on the two side chamber access tunnels and bore holes can begin.  This work will be done in parallel with the construction of the dome of the central chamber.  Thus, once the central chamber dome is finished, about one year after beginning of construction, we will be ready to begin the domes on the two side chambers. This parallel construction mode with the side chambers several months behind the central chamber allows efficient use of equipment and personnel while at the same time permitting us to test the construction plan on the central chamber before employing it on the two side chambers.  This overall construction procedure will result in one completed chamber in four years with the two side chambers available a few months later.

In year two, we will begin on the main, vertical cylindrical section of the central chamber.  There will be a series of 3 meter high cuts of a 50 meter diameter section, essentially the same size as the initial cut.  These cuts will made much faster since there is no need for vertical cable bolting.  Each of these cuts consists of three steps.  

The first step is to drill downward holes for explosives and then to set off these explosives.  By using a concentric circular layout and by staggering the firing of these explosives, it should be blast the rock toward the open central borehole region and minimize outward shock waves.  In addition, we plan to drill a series of vertical holes along the cylinder boundary to prevent outward transmission of shock waves and resultant fracturing of the cylinder walls.       

The second step is to move the blasted rock fragments into the two boreholes so that they can fall to the bottom.  Front end loaders in the bottom tunnel will move the broken rock to the rock dump near the shaft for hoisting.

The third step is to install horizontal, radial 60 ft long cable bolts into the cylinrical rock walls and immediately thereafter install the precast concrete liner sections.   These liner sections will be bolted into the rock and will dovetail into each other.  The joints between liner sections will be grouted so that the inner surface is smooth. Finally, we will spray on a layer of Mine Guard, a polyurethane coating that will serve as a inhibitor of radon flow from the walls and simultaneously a very smooth and low friction surface for the water tight plastic liner that will be installed at the end.

Each 3 meter high cut will take 10 weeks, 3 weeks for the blast hole drilling, 3 1/2 weeks for the rock removal and another 3 1/2 weeks for cable bolt installation.  The concrete liner installation will overlap the cable bolt and next layer blast hole drilling.  By staggering these activities by about 3 weeks between each of the three chambers, we can utilize three specialized crews, one for each of these tasks, and move them from one chamber to the next.  In addition, since separate equipment is used for each of these three tasks, we efficiently utilize the equipment and avoid having it stand idle for 2/3 of the time.

In order to permit horizontal entrance into the excavation region for each of the three meter high cuts, we will build a spiral ramp around each chamber.  Each cut will consist of a horizontal entry into the chamber region from this ramp.  Once a given layer is completed, the entry cut will be sealed with a concrete plug.  These ramps will permit machinery to easily move from one chamber to another or to the maintenance area and will provide a path for moving some of the blasted rock to the ouside bore hole.  At the end of the chamber construction process, these spiral ramps will serve as secondary ventilation links between the input and exhaust air systems and provide convenient local access to the bottom of each chamber.  For example, we could carry out water purification by withdrawing water from the bottom of the detector, pumping it up through a pipe in the spiral ramp to a purification system at the top and then returning it to the top of the chamber.   

The chamber will have an outer walkway that will go around the top of the chamber and a central gridwork that will be suspended from the upper dome cable bolts.  Both of these will be installed at the time the top cylinder cut is made.  The walkway will provide access to the upper ends of the side wall photomultiplier arrays.  The top gridwork will provide a mounting platform for the top layer of photomultiplier tubes as well as access for various calibration equipment. 

The final chamber construction task will be the installation of the inner plastic liners.  The conventional procedure is to use two liners, one a geotextile that serves as a cushion and also an inner fused water tight liner  Each of these materials are supplied on rolls.  We plan to mount the end of each roll against the rock along the upper walkway and then gradually lower the roller and so unroll the material along the vertical side of the cylinder.  The geotextile layers will be unrolled so that there is a butt joint between adjacent layers.  A tape seal will then be applied to these joints.  Access will be via a bosun chair suspension from the walkway.  Once the geotextile layer is complete, the waterproof layer will be lowered.  In this case, we will use an overlap joint which will then be heat sealed, again via the bosun chair suspension. 

The plastic liners along the bottom will be extensions of the side liners.  Before being brought into the mine, the bottom section of each roll will be cut into a triangular shape to conform to the bottom.  We may add an additional bottom plastic overlay to insure that this section is truly water tight.  It is worth noting that this multi-layer water tight construction is conventionally used to seal the bottoms of reservoirs and other bodies of water.

\subsection{Construction Timetable and Cost}
\label{twotimetable}

Mark Laurenti has worked out a detailed timetable and budget for the construction of these chambers including initial rock evaluation coring, construction of both top and bottom access tunnels, removal of waste rock, maintenance of mining equipment, etc.

The excavation process consists of continuous repetition of three separate tasks (1) drilling and blasting of rock, (2) removal of the rock rubble, and (3) installation of rock and cable bolts to stabilize the freshly exposed rock walls. Each excavation cycle is about 10 weeks with 3 weeks for each of the above three steps. There is a  cost savings in excavating three chambers at the same time, with a three week phase shift between steps in each module. This arrangement permits each of the three specialized crews to move from one excavation to the next every three weeks or so and continue using the same equipment and carry out their specialized tasks. In contrast, using one crew to sequentially do three different tasks will result in idle equipment for 2/3 of the time and inefficiency as they switch from one task to another.

For the three chamber mode, the cost of excavating each chamber is \$14.7M. This includes \$3.25M for the concrete liner and a 15\% contingency. In contrast, the cost of excavating a single chamber is \$16.9M including liner and contingency. The total required equipment cost is the same for both of these construction modes.

Assuming three shifts/day and 5 days/week operation, it will take 208 weeks or 4 years to excavate each 3 chamber group. This time could be reduced by going to a 6 or 7 day week. The Homestake Company frequently operated on a 6 or 7 day per week basis.

\subsection{Rock Removal}

Each kiloton chamber ($10^5 m^3$) involved the removal of about 419,600 tons of rock including access tunnels, domed roof, etc. For three chambers this results in 1,258,800 tons of rock in 4 years or 314,700 tons of rock per year. Since the hoisting capacity of the Wince 6-Ross shaft system is 750,000 tons per year, the simultaneous construction of three chambers utilizes only 40\% of the capacity of this shaft system.

\subsection{Equipment Cost}

Since all mining equipment has now been removed, new mining equipment will have to be purchased or leased. The required equipment, one Face Drill, two LHD loaders, 2 Bolters, 2 Underground Support Vehicles, 2 Lift Trucks, 1 LH Drill and 2 ITH Drills, costs about \$4.2M. It may be possible to arrange for leases instead of purchasing these items. Normal equipment maintenance has been included in the construction cost. The cost of this equipment
should probably be assigned to this specific task; it will be used full time during the construction period.

\subsection{Choice of Depth and Depth Dependent Cost}

There has been considerable discussion of depth necessary for very large detectors and the costs associated with deep detector locations. It is clear that the deeper the detector, the lower the cosmic ray muon and associated particle background. It is always preferable to have lower background. We can quantify the background limit by specifying that there be less than one cosmic ray related event per year with the megaton detector during the time that the accelerator beam is on. If we assume the accelerator beam is on for one microsecond per second, this requirement specifies an upper limit of $1.6 \times 10^{-6}$ ${\rm muons}/m^2$ sec., essentially the cosmic ray flux at about 7000 ft depth. The effect of this specification is that every event observed in the detector during the beam-on time is due to a neutrino from the accelerator without any cuts
whatsoever.

The question then is one of access and rock strength, namely, does a specific facility have ready access to a deep location and is the local rock structure capable of supporting large chambers. For Homestake the answer to both these questions
is {\bf Yes}. The present mine extends to 8000 ft, about 1000 ft deeper than the proposed detector location, and the rock seems strong enough to readily permit the excavation of large chambers.

At the end of section~\ref{lessons} we provide a comparison of costs of building the megaton Cerenkov detector at 6950 ft depth vs. at the 4850 ft depth. As indicated there, the maximum additional cost for putting the megaton Cerenkov array at 6950 ft vs. at 4850 ft is 5--6\% of excavation cost or less than 2\% of total detector cost.

\subsection{What Lessons About Depth Can Be Learned From Previous Experience?}
\label{lessons}

Detectors are located underground to reduce background in the detector due to cosmic rays. The deeper the detector, the lower the cosmic ray background. We have yet to have a detector that claimed to be ``too deep''. The only 
issues are: (1) is there a substantial additional cost associated with depth, and (2) are there technical limits associated with rock strength, etc. that limit depth at a given location? For many existing laboratories, depth is specified by what is available at that facility. Only two locations, the Sudbury mine, SNO, and the Homestake Mine, chlorine radiochemical experiment \cite{davis}, have multiple levels available. SNO chose to be at 6800 ft, essentially the same as the proposed megaton detector. Since chlorine was the first underground neutrino detector, there were no precedents and so it might be instructive to review the sequence of events that led to its location.

In 1962, Ray Davis tested a small perchloroethylene detector in a limestone mine in Barberton, Ohio at a depth of 2200 ft. The $^{37}Ar$ production was completely dominated by cosmic rays. That started  a search for a much deeper site. There were two possibilities in the U.S., with Homestake the preferable one. At that time, in 1965, 4850 ft was the deepest level that the Company would agree to. At the time the prediction for the solar neutrino signal was larger than now, there was  no thought about signal depression because of neutrino flavor conversion and no one expected a final measurement with a 5\% statistical precision. By the early to mid-1970's, it was already clear to us that the cosmic ray induced background was too large, given the observed signal, and that we needed a larger and deeper detector. At the time, the Company was not willing to consider a deeper,larger detector.

The final result was that the cosmic ray induced signal is 10\% of the solar neutrino signal in the chlorine detector and the uncertainty in that background signal is the largest contributor to the systematic uncertainty. The lesson is clear---locate detectors as deep as possible and be sure that there is a roadmap to detector enlargement.

There are two depth dependent costs, the cost of hoisting rock and the cost of rock and cable bolting. To estimate this effect, we determine the difference in costs between identical chambers built at the 4850 ft level (the bottom level of the Ross shaft, the upper hoist system, and the beginning of the Winze 6, the lower hoist system) and the 6950 ft level. The direct manpower costs
 for hoisting the extra 700 meters in the Wince 6 are about \$0.30/ton. The power costs add another \$0.20/ton for a total of \$0.50/ton or \$210,000 per 
100 kiloton module, where shaft maintenance costs have not been included.

The incremental rock support costs are more difficult to determine. The cable bolting planned and budgeted for these modules is far greater than required. This was done to insure that the chambers would have a minimum 50 year occupancy. A similar approach to corresponding excavations at the 4850 ft level might result in exactly the same bolting difference in rock stress between the two
 levels. The rock stress in the Homestake and Poorman formations, the formations that have been extensively studied in the mine, are rather surprising. The measured vertical stress $S_v = 28.3 \times {\rm D kPa}$, where $D$ is the depth in 
horizontal stress is very direction dependent. Along the high stress
horizontal axis $S_{h1} = 14,328 + 12.4 \times {\rm D kPa}$, while along the low stress horizontal axis, $S_{h2} = 834 + 12 \times {\rm D kPa}$. Presumably the high horizontal stress results from the rock folding that resulted in the upbringing of the gold ore deposit to the surface and thus to its discovery.

We assume that the effective stress at 6950 ft is about 35\% greater than the corresponding stress at the 4850 ft level. Since the total cost of the cable and rock bolts is \$910,000 and the related labor, including benefits, is about the same, we assign a depth dependent cost increase of \$630,000 for rock support. Combining this with the increase in hoisting costs gives a total of \$838,000 or 6\% of the total construction cost. Note that this is less than 2\% of the complete detector cost.

However, there are three offsetting costs that reduce the cost of constructing
the Cerenkov detector array at 6950 ft. vs. at 4850 ft. The first of these is the water fill. The total water fill for the megaton detector is 250 million gallons. Removing that much water from the local streams would be quite significant, especially given the present drought conditions in the area. Instead, we plan to use the water that is being pumped from the bottom of the mine at the 8000 ft. level. This water will be purified to remove any light scattering or absorbing material and any radioactive contaminants. Since the mine now pumps out about 350 gallons per minute, it will require about 1.4 years worth of water distributed over the construction time of the entire detector. For a detector at the 6950 ft. level, this water is only pumped up 1000 ft. while for a detector at 4850 ft., the water must be pumped up about 3100 ft. The cost savings here is about 1/4 of the increase in rock hoist cost or about \$50,000. The second offsetting cost is that of cooling the Cerenkov
detector. Operating the detector at $10^0$C gives 1/4 the photomultiplier noise of operation at $20^0$C. Since the rock temperature at the 4850 ft. level is over $35^0$C and still higher at 6950 ft., cooling will be necessary at either depth. The mine has an enormous refrigeration plant (24000 ton capacity) at the 6950 ft. level, with a fairly short path for the coolant from the refrigeration plant to the detector. A detector at the 4850 ft. level will either require a new refrigeration plant at that level or the installation of 2000 ft. of vertical coolant piping in the mine shaft. We have not estimated the cost of either of these steps, but they are clearly  substantial.

The third offsetting cost is that the level structure at 4850 ft. does not readily lend itself to the construction and ventilation system described above. If the upper detector access is a 4850 ft., then the lower, rock removal tunnel is at 5000 ft. There is no ventilation exhaust system at that level and waste rock would have to be raised in order to get it into the hoist system. The alternate approach, putting the top access at 4700 ft., would require additional
 excavation in order to provide the necessary tunnels for the upper access.

\subsection{Summary of Section~\ref{summ_const}}

A detailed construction plan for the construction of three 100 kiloton chambers in four years at the 7000 ft. depth in the Homestake Mine has been developed. The total construction cost of the three chambers is about \$44M or\$11M/year, as shown in detail in section~\ref{onemega}. In addition, there must be a one time purchase of about \$4.2M worth of mechanized mining equipment. The lead time in delivery of the mining equipment can be used to carry out coring of the rock region in which the detector array is to be
constructed.

In the next section we turn to the details of  construction and cost of one chamber of the megachamber in the absence of the detector that fills the chamber. 
A single chamber 50 m in diam $\times$ 50 m high is a taxing mining engineering accomplishment. We concentrate on a single one without detector components, which are discussed later, because we want to keep separate in our planning the large scale mining operations---a 50 meter high structure (even without dome) is the rough equivalent of a twelve story building---which are less familiar to us.

\section{Design and Cost of One Chamber of the
 Megachamber
and Projected Cost of
 Multiple Chambers}
\label{onemega}

The structural studies of large cylindrical excavations indicate that cylinders of 60 meter diameter by 120 meter length can be safely constructed at depths of 6800 to 7100 ft. in the Homestake Mine. Our goal is to construct a detector with a total mass of a megaton, which can be best achieved by constructing 10 separate 50 meter diameter by 50 meter high (100 kiloton) cylindrical chambers. In this section we consider the plans and the cost of constructing one such
 chamber, a 5 chamber system,  and  the 10 module megachamber system.

\subsection{Single Chamber Construction Costs}
\vspace{-1ex}

A detailed construction timetable and cost for the construction of these chambers was developed by Mark Laurenti.  It is worth repeating Laurenti's credentials, the Chief Mine Engineer for Homestake until the end of 2001, when the Company ceased all underground mining.  The development of this timetable and budget utilized his experience in budgeting and planning underground excavation and construction at the Homestake Mine.  Table~\ref{laurenti} summarizes the costs for the construction of a single 100,000 ton chamber.

\begin{table}[ht]
\begin{center}
\caption{\label{laurenti}Summary of Estimated Total Costs (M. Laurenti)}
\begin{tabular}{|l|r|} \hline \hline
Total Labor \& Benefits \phantom{xxxxxxxx} & \$5,506,656\\ \hline
Total Equipment Operation & 1,297,210\\ \hline
Total Supplies & 7,760, 490\\ \hline
Total Contractors & 123,000\\ \hline
15\% Contingency & 2,203, 103\\ [1ex]\hline \hline 
 Grand Total & \$16,890,460\\ \hline
\end{tabular}
\end{center}
\end{table}

\vspace{-.3in}


\input{tabletwo}

\subsection{Three Chamber Construction Cost}

As  pointed out above, the chamber excavation procedure consists of three approximately equal duration tasks, blast hole drilling and blasting, blasted rock removal and installation of rock support cables and concrete liner.  Each of these tasks requires specialized equipment and specific work skills.  In the construction of a single chamber, equipment is idle 2/3 of the time and the mine crew switches from one task to the next every three weeks or so.  Both of these situations lead to inefficiency in chamber construction and so higher construction costs.  There is an additional factor, the ratio of supervisory and equipment maintenance personnel to miners.  

By constructing three chambers at once with a phase shift between operations in each chamber, we can avoid these inefficiencies.  Now, we would have three separate crews, one for each of the above sub-tasks.  After the blasting crew drills the blast holes and blasts the next rock layer in the first chamber, they would immediately move on to the second chamber and do the same there.  At the same time the rock removal crew would move into the first chamber.  About three weeks later, the blast crew moves to the third chamber, the rock removal crew to the second chamber and the wall support crew into the first chamber.  Each crew would become  expert in its specific tasks and might be able to reduce the time required to carry them out.

{\bf Adding the non-concrete liner costs and the transport cost to the above
subtotal gives \$17,361,460 or \$41.40/ton excavated.}

We tested the validity of these cost estimates against several other measures. One of these involved a cost estimate that Dynatec, a highly regarded Canadian mine consulting and mine construction firm made in February 2001 as part of a laboratory development program. Although Dynatec used a somewhat different cost construction, the final Dynatec costs were in reasonable agreement with the Laurenti cost estimate.





\subsection{Historic Costs of the Homestake Mining Company}
 
Another way to establish the cost of excavation is to use the historic cost that the Homestake Mining Company published in its annual reports. For 1997, HMC gave \$49.88/ton; for 1998, HMC gave \$34.40/ton; and for 1999, HMC gave \$44.48/ton; an average of \$42.90/ton. All of these figures are per ton of ore extracted and processed to extract gold. The costs include all mine operating costs and exclude only taxes, depreciation and central corporate charges. Thus, these figures include the cost of gold extraction from the ore, charges for mine operation
such as ventilation, general mine maintenance, mine operation overhead, etc. The costs also include charges for exploratory work in looking for additional gold deposits, the removal of wate rock required to access the gold ore, etc. A year in which there was little exploratory work or waste rock removal would have a low cost per ton while one in which there was significant such effort would have a higher cost per ton.


Given these considerations, the historic HMC costs are in agreement with the summary of costs given above, well within the uncertainties of both numbers.

\subsection{Final Total for  Cost of Construction of the First Chamber.}

To the best of our understanding, the total cost of \$17,362,000  for the first chamber---including a 15\% contingency---is well-documented as indicated by
 Tables~\ref{laurenti} and \ref{minecosts}.  The detector components and the funding rate to produce a complete 3M module will be extra and we turn to them in section~\ref{onemodule}.

\subsection{Costs and Schedules for Multi-Chambers}


\subsubsection{Three Chambers}

As stated above, we plan to finish three chambers four to five years after the initial funding becomes available. This can be done either by starting to excavate three chambers at the same time in the first year, or somewhat more conservatively by carrying out excavation of a single chamber during the first year to gain experience, and proceeding to the second and third chambers as well as continuing with the first during the subsequent three to four years. This decision will be made when the opinion of the chief mining engineer at the time is available. In either case we anticipate a yearly expenditure rate of approximately \$15M for the first five years of operation which includes \$4.2M for the purchase of the required mining equipment. See Table~\ref{comparison} which compares single and three chamber construction costs.
\begin{table}
\begin{center}
\caption{\label{comparison}Comparison of single chamber versus three chamber cost}
\begin{tabular}{|l|l|r|r|} \hline \hline
Estimated Costs (\$MM) & & & \\ \hline \hline
& \# Of Chambers & 1\ \   & 3 \ \ \\ \hline \hline 
Labor \& Benefits & & \$5.51 & \$10.94\\ \hline 
Mining and Construction  & & &\\ \hline 
\ \ \ \ Equipment Operation & & \$1.30 & \$3.89\\ \hline 
\ \ \ \ Supplies & & \$4.51 & \$13.35\\ \hline 
\ \ \ \ Precast Concrete Liner & & \$3.25 & \$9.75\\ \hline 
Other (Outside Contractors) & & \$0.17 & \$0.52\\ \hline 
15\% Contingency & & \$2.20 &  \$5.74\\  [1ex] \hline \hline 
& TOTAL & \$16.94 & \$44.19\\ \hline
\end{tabular}
\end{center}
\end{table}

\subsubsection{Five Chambers}

The next three chambers can be started, all at the same time, during the final year of work on the first three chambers.
With this excavation schedule, it is likely that installation of a Cherenkov detector in at least one chamber can begin four years after the start of the excavation process, with other installations following soon after, depending on the flow of funds and photomultiplier tubes (PMTs). The feasibility of doing much of the work to achieve completed physics modules is the result of the layout of the individual modules shown in Fig.~\ref{f:array}, in which the spacing between adjacent modules,
450 ft., is sufficient to allow excavation of one chamber at the same time as work on liners or structures to support PMTs is carried out in other chambers.

In any event, we expect to have at least three, and perhaps as many as five,
modules ready for experiment at the end of seven years after the start of the program. See also section~\ref{theway}, {\bf The Way to the Megaton}.

\section{Design and Cost of One Module of the Cherenkov Detector}
\label{onemodule}

 At Penn, we have considerable experience in designing 
and constructing the detection systems of large neutrino detectors: in 1978, at 
BNL, a system with 13,000 proportional drift tubes (PDT) for tracking and 4,000 
PMT for calorimetry and data analysis \cite{BNL734}; in 1986 for the Kamiokande-II detector in 
Japan, the electronics and DAQ for a system of 1,000 20'' PMT \cite{what}; and in 1997 for 
the Sudbury Neutrino Observatory, a system of 9,500 PMT and  DAQ \cite{SNO}.

The Cherenkov detector to fill the chamber is 
composed of detecting
elements---the photomultiplier tubes (PMTs); the PMT reflectors;
the PMT supporting mechanical 
structures; the associated electronic and DAQ
systems; the calibration systems; and the water purification and cooling
system. Here we outline the  technologies and show preliminary
cost estimates for each of these systems.
It is important to note that all of these technologies are well 
understood, there are no new or unusual developments required, and
the draft designs and costs are based upon experience with successful
prior  detectors.

\subsection{Photomultipliers}
The core of an imaging Cherenkov detector is the array of PMTs.
 The complexity and cost of the detector depend
largely on the total number of PMTs while the
the total photocathode coverage determines  
energy resolution as well as the energy threshold and the timing
performance of the PMTs largely determines the reconstruction
accuracy. Therefore PMT selection is critical to the affordability
and to the sensitivity of the final detectors. Clearly much work needs to
be invested in the many details of actual PMT procurement, nevertheless it
is possible to make a reasonable estimate based upon experience and
recent contacts with possible manufacturers. 

Because the cost per unit area of photocathode declines somewhat with
larger PMTs and the cost of many of the supporting systems like cables
and electronic channels declines directly with smaller numbers of 
PMTs, it is sensible to think of relatively large diameter PMTs.
The very largest devices, the Hamamatsu 50 cm tubes, have, however,
a significant mechanical disadvantage in terms of stored energy vs.
envelope strength as demonstrated in the SuperKamiokande accident.
Therefore we have chosen to base our estimates on tubes of about
25 cm diameter. Hamamatsu, Electron Tubes, and Photonis all have
tubes in this class, all with specified performance superior to  
the tubes used in the previous generation of large water Cherenkov
detectors. For specificity, we have chosen the Photonis 27 cm
XP1804 for this estimate. The XP1804 claims a quantum efficiency
of 24\% at 420nm and a Transit Time Spread of 2.4ns FWHM. Recent
measurements of prototype tubes at Penn support these numbers.

For this estimate we use a total photocathode coverage of 10\%
which implies  15,500 PMTs per 100 kT module. The cost
used in the total PMT estimate is scaled from the 20 cm tubes used in SNO and
is  not based on manufacturer's quotations.  Nevertheless, all manufacturers 
have accepted, in principle, that the estimate of \$800 per PMT
is realistic. To this \$800 number we add the PMT testing costs plus
the cost of the base and the waterproofing that surrounds the base.

We make the further assumption that at the planned depth of the chamber it will
not be necessary to have a sophisticated anti-counter (see Figure~\ref{f:muons}
in section~\ref{intro}) and  have allowed only for a few meters of water
at the top and bottom of the cylinder to be optically isolated
and equipped with a small number of anticounter PMTs looking
up or down. This is similar to the SNO approach where only 91 PMTs 
are used to observe the outer light water volume.

\subsection{Reflectors}

Because effective photocathode coverage is important to the
resolution and sensitivity of an imaging Cherenkov counter, it is
efficient to include optical reflectors (or concentrators) on each
PMT in order to increase inexpensively  the effective area. As described
in  detail in reference \cite{cones}  it is
possible to achieve nearly a 75\% improvement in effective photocathode
area using well designed concentrators. Because the 3M modules need
somewhat larger angular coverage than SNO wished for its central
volume of $D_2O$ and because there is some corrosion seen after five
or more years, we have derated the improvement in coverage to 40\% for this
estimate. It is still cost effective to add reflectors as the reflector cost
per PMT is less than \$60, scaled from the SNO costs. There is also
a need for some design and development work to get to a production
design for the 3M detector.

\subsection{Mechanical Structure}

The individual PMTs must be fixed in well known locations evenly 
distributed across the interior surface of the 50 m by 50 m cavity.
To do this, a mechanical structure must be designed and
installed.   There are a number of suggested
schemes for achieving this mounting, but not yet any engineered
plan. Therefore we  estimate this system cost on   general
grounds. The SNO mechanical mounting was for a spherical surface,
much more complex than the simple cylinder of this detector, and therefore
not a good cost model. The SK scheme is plausible for this case, but we
do not have access to their cost data. 
Including installation labor, a simple model gives about \$150 per
PMT total cost for the structure. 

\subsection{Electronics and DAQ}

The basic SNO electronics system has performed nicely and that architecture
should work well for the proposed 3M detector. However, the custom integrated
circuits were manufactured in processes that are no longer available
and many of the commercial parts are obsolete or obsolescent so a
redesign in detail will be necessary. This will also present an
opportunity to remove some of the minor flaws of the SNO
implementation. 

The SNO production and development costs are well documented and  
we have  scaled  for inflation to arrive at a per PMT
cost of about \$108 for front end electronics; trigger system; low
voltage power; and high voltage power. The development costs are taken
as the same number of people-years  required for the original SNO system
plus silicon chip and printed circuit prototyping costs. 

The DAQ system involves largely development, but there is a small cost
for the computers needed for data recording; on line monitoring, for slow
controls (largely environmental controls); and miscellaneous networking
and GPS interfaces. This adds a fixed cost per module of about \$75k.
An additional central computing farm to support off line analysis would
probably be included in a full design report but is only mentioned
here. The DAQ development costs are assumed similar to SNO's at about
five people-years.

\subsection{PMT Cables and Connectors}

The cable running from each PMT to the electronics, presumably located
around the periphery of the module above the water surface, is
 a significant cost item. For a 45 m x 45 m detector and equal
length cables to allow an isochronous trigger, each cable needs to
be at least 70 m long. High quality waterproof transmission line
similar to that used in SNO costs about \$0.50 per meter to which one
must add the assembly and testing costs and the connectors. 

It is possible to imagine for the right circular cylinder geometry of
a 3M module that it would be possible to eliminate a {\em wet end}
connector and simply pot the cable and base as a unit. For conservatism,
and because there are unknown installation costs associated with pre-
attached cables, we have assumed a {\em wet end} connector with full
waterproofing for a total assembled cable cost of about \$70.

\subsection{Calibration Systems}

A number of levels of calibration are required of an imaging
water Cherenkov detector. The most basic level is simply a series
of electronic calibrations of charge and time 
in well defined units, which  is included in
the electronics costs above. However, it is also necessary to have an
optical calibration system to define a common time reference and to
measure the optical properties of the detector. We choose a fast,
multi-wavelength laser source and a diffusing ball capable of being
moved
through a large fraction of the total detector volume. Finally, one
needs  handles on actual energy scales. The  peak at the end point of 
stopped muons is 53 MeV.
At least one other energy point is needed to give  the
slope of the energy calibration curve. It is 
 useful to consider the SK scheme of a small electron accelerator
to provide a  versatile calibration source---this is especially
attractive if it can be shared amongst all the modules.
 We do include a laser system;
a source positioning apparatus; and a neutron generator and decay chamber
for one nuclear source as a development cost to arrive at a total
of about \$400k for calibration systems.

\subsection{Water Purification and Cooling}

The water system is modeled on the SNO light water
system which has produced water of exceptional cleanliness and
low radioactivity. For costing purposes we have simply scaled the
SNO costs by one half the volume ratio on the assumption that we would
require the same level of transparency as SNO - greater than 200 m in the
region between 350 and 450 nm (very near the Rayleigh limit). This
implies that we would want to have the same level of sophisticated
reverse osmosis and filtering equipment 
and  would want to maintain nearly the same rate of turnover.
However, the full volume ratio is clearly an overestimate as fourteen
 three inch pipes cost much more than 
one twelve inch pipe and fourteen small pumps are more expensive than
one or two much larger pumps. 

In addition to the purification system, we also require the water to
be degassed to inhibit biological activity and cooled to further inhibit
biology and to keep the PMT noise rates low. The degasser is included in
the SNO costs as are heat exchangers. The Homestake Mine already has a
large cooling plant and so within the cost of plumbing from the 
existing chiller to the detectors that cost is covered. As the SNO
costs also include plumbing from a (closer) chiller station, we have
simply scaled those costs by our factor of seven (one half of the volume
ratio of 14). A full design effort will be able to itemize and
probably reduce these estimates.

\subsection{Summary of Costs}

\paragraph{Development Costs} for the major detector systems are
extrapolated largely from the SNO cost history and are summarized in 
Table~\ref{developmentcosts}. These costs are shared over all
ten 3M modules. Some of these tasks require professional engineering
while some (e.g. DAQ systems) are largely physicist tasks.

\paragraph{Procurement Costs} for module  hardware see Table~\ref{procurementcosts}. Note that the
cost of the electron accelerator for calibration is shared amongst all
ten 3M Modules.

\begin{table}
\caption{\label{developmentcosts}Estimated development costs by subsystem.}
\begin{tabular}{|l|r|l|}
 \hline \hline
Subsystem & Cost, k\$ & Comments\\ \hline \hline
Photomultipliers & 230 & PMT tests; base; waterproofing \\ \hline
Reflectors & 120 & Material selection and testing \\ \hline
Mechanical & 275 & Engineering and prototypes \\
Structure & & \\ \hline
Electronics & 1,450 & Chip, board, system design and prototype \\ \hline
DAQ & 250 & Programming; co-developed with Electronics \\ \hline
PMT Cables & 55 &Cable and connector specification \\ \hline
Calibration & 250 & Design of manipulators plus specification of sources \\ 
\hline
Water Purification & 350 & Engineering \\ \hline \hline
Total all Modules &  { \bf 2,980 }   & \\ \hline
\end{tabular}
\end{table}

\begin{table}
\caption{\label{procurementcosts}Estimated procurement costs by subsystem.}
\begin{tabular}{|l|r|l|}
 \hline \hline
Subsystem & Cost, k\$ & Comments\\ \hline \hline
DAQ & 75 & Computer and Networking hardware \\ \hline
 Calibration & 300 & Laser, manipulator, source(s), $e^-$ generator.  \\ \hline
Water Purification & 4,500 & M\&S plus labor \\ \hline \hline
Total per Module &  {\bf 4,875 }     & \\ \hline
\end{tabular}
\end{table}

\paragraph{Production Costs} for the items directly in the photomultiplier
chain and therefore directly proportional to the number of PMTs 
installed are shown in Table~\ref{productioncosts}. These estimates include
assembly and quality assurance labor as well as final installation labor 
(under Mechanical Structure).

\begin{table}
\caption{\label{productioncosts}Estimated production costs per installed PMT.}
\begin{tabular}{|l|r|l|}
 \hline \hline
Subsystem & Cost, \$ & Comments\\ \hline \hline
Photomultipliers & 850 & PMT; base; waterproofing \\ \hline
Reflectors & 60 &   \\ \hline
Mechanical & 150 & Materials plus labor \\
Structure & & \\ \hline
Electronics & 110 & Chip, board, system design and prototype \\ \hline
PMT Cables & 125 &Cable and connectors \\ \hline \hline
Total per PMT & {\bf 1,295} & \\ \hline \hline
Total per Module & {\bf 20,073 k\$  } & for 15,500 PMTs \\ \hline
\end{tabular}
\end{table}

{\bf Thus, for the first module, we have a total cost, including all
development costs, of about \$28 M. Applying a 25\% contingency
gives a first Module cost of about \$35 M.} These costs are dominated
by the per PMT production costs and so peak in the later years of the
construction. Using the cavity construction schedule above with
beneficial occupancy for detector components at the end of the fourth
year, we estimate a funding profile shown in Table~\ref{profile}.

\begin{table}
\caption{\label{profile}Detector costs by project year.}
\begin{tabular}{|l|r|l|}
 \hline \hline
Project Year & Cost, k\$ & Comments\\ \hline \hline
1 & 600 &  Design and Engineering, all systems \\ \hline
2 & 6,800 &   Begin prototyping most systems, order PMTs \\ \hline
3 & 9,000 &   Begin procure water system, test PMTs \\ \hline
4 & 9,400 &   Finish most procurement and testing \\ \hline
5 & 2,100 &   Install detector subsystems \\ \hline  \hline
Total per Module & {\bf 28,000  } &  \\ \hline
Total with Contingency & 35,000 & \\ \hline \hline
\end{tabular}
\end{table}

\section{Total Costs and Schedules of a Complete Single Component and 
Five Component Detector Array}
\label{totalcosts}

The estimated cost of a single cavity in section~\ref{onemega} when added to the estimated
cost of the PMTs and support electronics and structures in 
section~\ref{onemodule}  yields a total cost of \$52.4M  including a 15\% contingency on chamber production and 25\% contingency on detector production and installation. With what we know now, we estimate less than \$300M for the five module array.

\section{The Way to the Megaton}
\label{theway}

\subsection{Construction on the Way to the Megaton}

The question of at what rate it will be possible to proceed to the full operational array depends on factors that are too uncertain at this time to allow a prediction beyond an extrapolation from the rate observed in production of the first half of the array. We are unable now to determine what will be the limit to progress; e.g.,
the rate of procurement of PMTs or new types of photosensitive devices which will require new detector designs, etc. 

Nevertheless, as long as present estimates are borne out by future experience, we visualize producing one completed detector module on average every year, so that the megaton array will be complete and operational in not more than 12 years, and cost less than \$600M. We emphasize, however, that the 3M detector will begin acquiring useful data continuously as modules become active. Completion of the first half of the array will allow the search for CP-invariance violation to be carried out with the high statistics samples as described in section~\ref{CP}. Note again that construction and data-taking can be conducted simultaneously.

\subsection{Physics on the Way to the Megaton}

One of the virtues of the 3M detector is that it will begin doing valuable experiments {\it as soon as each of its detector modules is operational.}
An important case in point is as follows.

\subsubsection{Conclusive Demonstration of Neutrino Oscillations and Measurements of $\Delta m^2_{12}$; Measurements with Extra-Terrestrial Neutrinos}

In section~\ref{intro} we pointed out the unique physics yield of a very long base line neutrino oscillation experiment employing a modest energy neutrino beam; this includes a   broad range of $\Delta m^2$ that can be measured and the multiple
 nulls that can be observed which demonstrate the oscillatory nature of the signal. This is discussed more fully in section~\ref{CP}. However, the five module array is not necessary for meaningful data to be taken. For the upper part of the $\Delta m^2$ range, where $\Delta m^2_{23}$ is expected to dominate, the signal of interest is the disappearance of $\nu_{\mu}$. This is what has been observed  for neutrinos originating in the atmosphere. The neutrino energy distribution of the AGS-Homestake beam has a longer flight path than any existing or proposed experiment, resulting in a much larger range of sensitivity for $\Delta m^2_{23}$. For example, if $\Delta m^2_{23} = 2 \times 10^{-3} \ {\rm eV^2}$, the $n=2$ null occurs at $E_{\nu} = 1.37$ GeV and the $n=3$ null occurs at $E_{\nu} = 0.85$ GeV. Both of these energies are within the natural distribution of the AGS neutrino beam.

In addition to the $\nu_{\mu}$ experiment discussed above, there is also the
 probability of $\nu_e$ appearance due to $\nu_2 \rightarrow \nu_1$
 transitions.  For $\Delta m^2_{12} = 6 \times 10^{-5} \ {\rm eV^2}$, the most recent value of $\Delta m^2$ solar,  a search for $\nu_e$ appearance will directly address the limiting value of $\Delta m^2_{12}$ even with a modest statistical sample of events.

Among the several attractions of this program at Homestake are that it can be started within five years after approval of this proposal so that testing of the apparatus and learning by the experimenters can begin early. Early measurements will be on extra-terrestrial neutrinos. 

Finally, neither the length of the construction period nor the anticipated cost of the 3M detector is large in relation to plans and expenditures now usual for major apparatus in high energy and elementary particle physics. That statement is based, for example, on the magnitude of efforts and expense that have rightly gone
into the search for CP-violation in the b-quark family alone in the past few years. Another example is the effort through LIGO to find direct evidence for gravitational waves. The Homestake Laboratory and the 3M detector are further examples that exploration of new frontiers of science can no longer be implemented by improvisation at small cost.

\section{Outreach}
The topics of this proposal are of great interest to students both at high
 school and college levels,
as well as to the general public.  In the past experimental program at the
 Homestake mine, there has been participation by both high school students from the local Lead and
 Deadwood, South
Dakota communities and undergraduate students from the nearby South Dakota School of
 Mines and
Technology and the Black Hill State University.   In addition, a student at
 South Dakota State
University in Brookings, 400 miles from the Homestake Mine, did a senior
 thesis on the physics
of neutrinos as they related to the ongoing experiment.  We intend to
 continue and extend this aspect of
outreach with the proposed program. 

The physicists at the above schools and at the University of South Dakota
 have been invited to
participate in this program and have been encouraged to promote active
 participation by their
undergraduate students. To encourage active  participation
 and to provide a
view of this science for the general  population, we  have also
 given talks throughout the state to various professional and civic groups.  This effort will continue with the proposed program.

The anticipated establishment of the National Underground Science and Engineering Laboratory
 at the
Homestake Mine has greatly increased the awareness of and interest in
 science throughout
South Dakota.  The large physical separations between the population centers
 in the eastern and
western parts of South Dakota somewhat limit frequent visits to the
 Laboratory by those in the
eastern half of the state. Fortunately, South Dakota has an extensive fiber
 optic network that
links all the K-12 schools of the state together.  We will utilize this network to
provide tutorials of our program for that school population as
 well as  for the
general population.  We will also use this link to provide
 ``live" views of the
detector assembly and operation  when that is appropriate.

 We will organize visits to the laboratory for high school science
 teachers to acquaint
them with the both the scientific ideas being tested and developed and to
 give them a clearer
picture of the apparatus involved.  These visits can then be followed with
 occasional live internet
``ask the physicist" sessions with their students.  It will be useful
 for NUSL to establish a
summer Science Academy for high school students that would involve some
 course material and
participation in one of the experimental programs at Homestake. 
 We would be
 extremely interested
in conducting such a program.

\end{document}

%% file: tabletwo.tex
\begin{table}
\caption{\label{minecosts}Breakdown of Estimated Total Costs in Table~\ref{laurenti}.}
\begin{small}
\begin{center}
\begin{tabular}{lrrrrr}
{\bf Description} &  {\bf \$/Unit} & {\bf Adj.\$/Unit} & {\bf Units} &  &  {\bf Total} \$\\
{\bf Equipment Operation}  & & & &   & \\
\multicolumn{6}{l}{(Operating cost includes all consumables such as tires, fuel, filters, parts and misc.)}\\
\ \  1 Face Drill       &  \$0.42 & \$0.50 &\phantom{xxxx} 90,200  Tons  & & \$ 45,461\\
\ \  4 LHD              &  \$0.90 & \$1.08 & 419,600 Tons  & & \$453,168\\
\ \  2 bolter           &  \$0.97 & \$1.16 & 419,600 Tons  & & \$473,762\\
\ \  2 UG Support Veh   &  \$0.10 & \$0.12 & 419,600 Tons  & & \$5,035\\
\ \  2 Lift trucks      &  \$0.10 & \$0.12 & 419,600 Tons  & & \$5,035\\
\ \  1 LH Drill         &  \$0.58 & \$0.70 & 419,600 Tons  & & \$169,384\\
\ \  2 ITH Drill        &  \$0.58 & \$0.70 & 360,100 Tons  & & \$145,365\\
 & & & & &  \rule{16mm}{.5mm}\\
 & & & &  & {\bf \$1, 297,210}\\
{\bf Supplies} & & & & &  \\
\ \ Blasting Supplies      &  \$0.86 & \$1.03 & 419,600 Tons  & & \$431,957\\
\ \ Drill Bits              &  \$0.60 & \$0.72 & 419,600 Tons  & & \$300,684\\
\ \ Drill Steel             &  \$0.25 & \$0.30 & 419,600 Tons  & & \$127,236\\
\ \ Gen'l Operating Supplies & \$1.47 & \$1.76 & 419,600 Tons  & & \$738,798\\
\ \ Cable Bolts              & \$4.00 & \$4.80 & 147,200 ft.  & & \$706,560\\
\ \ Rock Bolts               & \$0.57 & \$0.68 & 300,000 Tons  & &  \$204,236\\
\ \ Hoses                    & \$0.06 & \$0.07 & 419,600 Tons  & & \$28,298\\
\ \ Lubricants               & \$0.04 & \$0.05 & 419,600 Tons  & & \$19,937\\
\ \ Pipe                     & \$15.00 & \$18.00 & 3,070 ft.  & & \$55,260\\
\ \ Safety Supplies          & \$ 0.05 & \$0.06  & 419,600 Tons  & & \$26,345\\
\ \ Small Tools              & \$ 0.01 & \$0.01  & 419,600 Tons  & & \$6,122\\
\ \ Vent Bag                 & \$ 0.01 & \$0.02  & 419,600 Tons  & & \$7,365\\
\ \ Spot Coolers             & \$18,611.63 & \$22,334/ & 4 years &   & \$357,343\\
& & \ \ \ \ \ \ \ \  cooler/year & &  &\\
\ \ Shotcrete \& Screen      & \$130.00 & \$156.00 & 2,810 cuYd  & & \$438,360\\
\ \ Precast Cement Liners    & \$1,300.00 & \$1,300.00   & 2,499  & & \$3,248,700\\
\ \ Spray Liner              & \$5.00 & \$5.00 & 105,616sqft & &  \$528,080\\
\ \ Top Deck Material        & \$20.00 & \$20.00 & 21,124sqft & &  \$422,480\\
\ \ Shop Construction        & \$15.00 & \$18.00 & 3,416sqft & &  \$61,488\\
\ \ Diamond Drill            & \$7.00 & \$8.40 & 6,100ft & &  \$51,240\\
& & & & &   \rule{16mm}{.5mm}\\
& & & &    & {\bf \$7,760,490}\\
{\bf Outside Contractor} & & & & &  \\
\ \ Borehole                 & \$ 250.00 & \$300.00 & 410ft.  & & 123,000\\
 & & & & &  \rule{16mm}{.5mm}\\
 & & & &   & {\bf \$123,000}\\

\end{tabular}
\end{center}
\end{small}
\end{table}